\documentclass[twocolumn,showpacs,amsmath,amssymb,prd,floatfix]{revtex4}

\usepackage{graphicx}
\usepackage{dcolumn}
\usepackage{bm}
\usepackage[varg]{txfonts}

\begin{document}

\title{

Nonlinear perturbation theory with halo bias and redshift-space
distortions\\ via the Lagrangian picture


}

\author{Takahiko Matsubara}
\email{taka@a.phys.nagoya-u.ac.jp}
\affiliation{%
Department of Physics, Nagoya University,
Chikusa, Nagoya, 464-8602, Japan
}%

\date{\today}

\begin{abstract}
    The nonlinear perturbation theory of gravitational instability is
    extended to include effects of both biasing and redshift-space
    distortions, which are inevitable in predicting observable
    quantities in galaxy surveys. Weakly nonlinear effects in galaxy
    clustering on large scales recently attracted a great interest,
    since the precise determination of scales of baryon acoustic
    oscillations is crucial to investigate the nature of dark energy
    by galaxy surveys. We find that a local Lagrangian bias and
    redshift-space distortions are naturally incorporated in our
    formalism of perturbation theory with a resummation technique via
    the Lagrangian picture. Our formalism is applicable to any biasing
    scheme which is local in Lagrangian space, including the halo bias
    as a special case. Weakly nonlinear effects on halo clustering in
    redshift space are analytically given. We assume only a
    fundamental idea of the halo model: haloes form according to the
    extended Press--Schechter theory, and the spatial distributions
    are locally biased in Lagrangian space. There is no need for
    assuming the spherical collapse model to follow the dynamical
    evolution, which is additionally assumed in standard halo
    prescriptions. One-loop corrections to the power spectrum and
    correlation function of haloes in redshift space are explicitly
    derived and presented. Instead of relying on expensive numerical
    simulations, our approach provides an analytic way of
    investigating the weakly nonlinear effects, simultaneously
    including the nonlinear biasing and nonlinear redshift-space
    distortions. Nonlinearity introduces a weak scale dependence in
    the halo bias. The scale dependence is a smooth function in
    Fourier space, and the bias does not critically change the feature
    of baryon acoustic oscillations in the power spectrum. The same
    feature in the correlation function is less affected by nonlinear
    effects of biasing.
\end{abstract}

\pacs{
98.80.-k,
95.35.+d,
95.36.+x,
98.65.-r,
}
\maketitle

\section{\label{sec:intro}
Introduction
}

The nonlinear perturbation theory of gravitational instability has
recently attracted renewed interest. As precision measurements of the
large-scale structure of the universe become possible, theoretically
accurate modeling is essential to interpret the observational data. It
is recognized that the linear perturbation theory is not satisfactory
for this purpose. The importance of nonlinear perturbation theory
resides in the era of precision cosmology.

A strong motivation for developing the perturbation theory is to model
the baryon acoustic oscillations (BAOs) imprinted in the large-scale
structure. Acoustic waves which propagate in the baryon-photon plasma
of the early universe freeze out at the recombination epoch, and the
sound horizon at that epoch is imprinted in spatial fluctuations of
photons and baryons \cite{BAOtheory,CMBobs}. The BAOs provide a
standard ruler \cite{BAOSR} to geometrically investigate the expansion
history of the Universe \cite{AP79}. The dark energy component is
efficiently constrained by galaxy surveys of an intermediate- to
high-redshift universe \cite{GeomTest,MSDE}. Using the BAOs as a
standard ruler, large galaxy surveys are expected to provide a robust,
promising way of constraining the nature of dark energy
\cite{BAODE,Mat04}. Recent observations of BAOs in modern galaxy
surveys \cite{BAOLSS} prove the method works well.

Even though the BAO scale is quite large $\sim 100\,h^{-1}{\rm Mpc}$,
detailed structure of BAOs in galaxy clustering is affected by
gravitationally nonlinear evolution after the recombination epoch. The
BAO signature in the power spectrum or in the correlation function is
deformed by nonlinearity in a lower redshift universe where realistic
galaxy surveys are possible \cite{BAONL,BAONLRD,Ang05,ESW07,SBA08}.

In galaxy redshift surveys, the clustering of dark matter is not
directly observable. There are two major sources in the difference
between the clustering pattern of dark matter and that of galaxies:
galaxy biasing and redshift-space distortions. The spatial pattern of
galaxy distribution is not necessarily the same as that of dark
matter, and the galaxies are biased tracers of mass \cite{Bias, BBKS}.
The redshift of a galaxy does not purely reflect the Hubble flow, and
the Doppler shift by a peculiar velocity is inevitably added. Thus the
spatial pattern of clustering of galaxies is distorted in redshift
space \cite{RDist,Kaiser87}. In the linear regime, the power spectrum
of biased objects $P_{\rm obj}(k)$ is usually assumed to be
proportional to that of mass $P_{\rm m}(k)$ in real space: $P_{\rm
  obj}(k) = b^2 P_{\rm m}(k)$, and there is a theoretical reason,
which is known as ``the local bias theorem'' \cite{LocBias,SW98}.
Assuming linear dynamics, a linear bias, and a linear velocity field,
the power spectrum of biased objects in redshift space is given by
\cite{Kaiser87}
\begin{equation}
  P^{\rm (s)}_{\rm g}(\bm{k}) =
  b^2 \left(1 + \beta \mu^2 \right)^2
  P_{\rm m}(k),
\label{eq:0-1}
\end{equation}
where $b$ is the linear bias factor between galaxies and mass, $\beta
\simeq \Omega_{\rm m}^{0.6}/b$, and $P_{\rm m}(k)$ is the power
spectrum of mass in real space.

The linear formula of Eq.~(\ref{eq:0-1}) applies only when the
redshift is large enough, or when the wave number $k$ is small enough.
However, the linear formula in redshift space is not sufficiently
accurate for typical redshift surveys \cite{Sco04}. Extending the
linear formula of Eq.~(\ref{eq:0-1}) to include nonlinear effects is
far from trivial. A straightforward application of the nonlinear
perturbation theory \cite{SPT,SPTdiag,MSS,BCGS02} with a local biasing
scheme \cite{FG93} has been studied so far
\cite{HMV98,Taruya00,McD06,HaloPT}. In the local biasing scheme, the
number density of biased objects is a local function of the smoothed
density field of mass in Eulerian space. This scheme seems to work as
long as the tree-level perturbation theory is adopted \cite{SCF99}.
However, when loop corrections are considered, there appears a
conceptual problem in the local biasing scheme. It turns out that the
one-loop corrections, even on large scales, strongly depend on the
artificial smoothing scale \cite{HMV98}. In Ref.~\cite{McD06}, a way
to remove the dependence on the smoothing scale by renormalizing bias
parameters is proposed. While this could be a solution to
phenomenologically represent the power spectrum of biased objects
\cite{JK08}, it is not guaranteed that the procedure actually reflects
the physical nature of biasing, since the biasing is more or less a
nonlocal process.

It is true that the biasing is difficult to be included exactly in an
analytic framework, since the galaxy formation is a highly nonlinear
process. The next best thing is to find a good analytic model. In this
respect, a plausible model is provided by a halo approach
\cite{MW96,MJW97,SL99,ST99,Sel00,PS00,MF00,SSHJ01,CS02}, which is
based on the extended Press and Schechter (PS) model
\cite{PS74,ExtPS}. In this approach, the galaxy biasing is described
through two steps. In the first step, the formation and clustering of
dark matter haloes are analytically modeled by using the extended PS
model. The spherical collapse model is combined to include the
gravitational evolution of halo positions. In the second step, the
distribution of dark matter or galaxies within haloes is empirically
modeled with a number of assumptions
\cite{SB91,Sel00,PS00,MF00,SSHJ01,CS02}, which include complicated
physics such as gas cooling, star formation, feedback effects from
supernovae, and so forth \cite{SemiAna}.

While the second step involves many parameters which should be fitted
by observations or simulations, the first step is theoretically less
uncertain because of its purely gravitational nature. On large scales,
the second step is not important and the power spectrum of nonlinear
objects is approximately the same as that of haloes. Therefore,
determining the power spectrum of haloes is an important step toward
understanding the nonlinear structure formation and biasing. In the
usual halo approach, the halo clustering is modeled by linear dynamics
and linear bias factors \cite{MW96,Sel00,PS00,MF00,SSHJ01}. Recently,
an attempt to incorporate the halo bias with nonlinear perturbation
theory in a framework of local biasing scheme has been made
\cite{HaloPT}. There still remains conceptual arbitrariness in the
latter formulation, mainly because the halo bias is intrinsically
nonlocal in Eulerian space and does not fit well into the local
Eulerian biasing scheme. From the construction, the halo bias is local
in Lagrangian space \cite{MW96}, and is therefore nonlocal in Eulerian
space, because positions of mass and biased objects are displaced by
dynamical evolution.

In this paper, a natural approach to incorporate the halo bias and
redshift-space distortions with nonlinear perturbation theory is newly
developed. To account for the locality of halo bias in Lagrangian
space, our formulation is based on the Lagrangian perturbation theory
(LPT) \cite{LPT}, instead of the standard Eulerian perturbation theory
(EPT). In the original halo approach, the spherical collapse model is
adopted to take into account the dynamical evolution in Eulerian
space. However, the dynamical evolution of haloes is more naturally
described in the Lagrangian picture \cite{LHalo}. It is not
straightforward to obtain an analytically useful formalism which
combines the LPT and the halo approach. A main obstacle is that the
observable quantities reside in Eulerian space, while the calculations
in LPT directly give Lagrangian quantities. In paper I \cite{Mat08}, a
new approach is developed to overcome this point, partially expanding
the Lagrangian variables in Eulerian space. The resulting expression
contains an infinite series of perturbations in terms of the EPT, and
this approach offers a simplified technique of resumming cosmological
perturbations, such as done in renormalized perturbation theory and
its variants \cite{RPTvar, NewPT}. While our technique is not suitable
enough to describe the fully nonlinear regime which other
renormalization techniques aim at, we have shown in paper I that our
technique is accurate enough in the quasilinear regime, and most
importantly, predictions in redshift space are straightforward. To
date, our technique offers the only way of obtaining resummed power
spectra in redshift space. In this paper, we show that the local
Lagrangian bias, including the halo bias, is also straightforward to
be incorporated in the approach of paper I, on top of the
redshift-space distortions. The local Lagrangian biasing scheme is not
equivalent to the local Eulerian biasing scheme, and the conceptual
problem about the strong dependence on smoothing scales which appears
in the local Eulerian biasing scheme described above is not present in
our Lagrangian approach.

The rest of the paper is organized as follows. In
Sec.~\ref{sec:LLBias}, our basic formalism is described. This
formalism is applicable not only to the halo bias but also to a local
Lagrangian bias in general. Linear and one-loop results are presented.
In this general theory, we have parameters which are related to the
Lagrangian biasing scheme. In Sec.~\ref{sec:HaloLPT}, those Lagrangian
bias parameters are calculated from the fundamental concept of the
halo model. Effects of halo bias and redshift-space distortions on BAO
scales are demonstrated both in the power spectrum and in the
correlation function. Our conclusions are summarized in
Sec.~\ref{sec:concl}. In the Appendix, details of one-loop
calculations in our framework are outlined.

\section{\label{sec:LLBias}
Local Lagrangian bias
}

\subsection{\label{subsec:NLPS}
Nonlinear power spectrum with a local Lagrangian bias
}

In this section, we develop a method to track nonlinear evolution of
the Lagrangian bias. Our method in this section is not restricted to
the halo bias model, and is applicable to any bias defined by a local
function of linear density field in Lagrangian space.

In the Lagrangian approach to track dynamical evolution of
cosmological density fields, a set of each trajectory of a mass
element, $\bm{x}(\bm{q},t)$, where $\bm{q}$ are initial Lagrangian
coordinates, describes the whole property of the density field. A
displacement field $\bm{\Psi}(\bm{q},t)$ is defined by
\begin{equation}
  \bm{x}(\bm{q},t) = \bm{q} + \bm{\Psi}(\bm{q},t), 
\label{eq:1-1}
\end{equation}
and is considered as a fundamental variable of the mass density field.
Since the initial mass density field is sufficiently uniform, the
Eulerian mass density field $\rho_{\rm m}(\bm{x},t)$ at any given time
$t$ satisfies the continuity relation,
\begin{equation}
    \rho_{\rm m}(\bm{x},t) \,d^3x = \bar{\rho}_{\rm m} \,d^3q,
\label{eq:1-2}
\end{equation}
where $\bar{\rho}_{\rm m}$ is the global mean density of mass.

On the other hand, the fluid elements in which biased objects such as
haloes reside are not uniformly distributed in Lagrangian space.
Therefore, the continuity relation between the Eulerian density field
of the biased objects, $\rho_{\rm obj}^{\rm E}(\bm{x},t)$, and
corresponding Lagrangian density field, $\rho_{\rm obj}^{\rm
  L}(\bm{q})$, is given by
\begin{equation}
  \rho^{\rm E}_{\rm obj}(\bm{x},t)\,d^3x
  = \rho^{\rm L}_{\rm obj}(\bm{q})\,d^3q.
\label{eq:1-3}
\end{equation}
The density field $\rho^{\rm L}_{\rm obj}(\bm{q})$ represents the
initial distribution of locations where biased objects form later.
Since the formation of nonlinear structure is too complex to be
analytically described from the first principle, the initial density
field $\rho^{\rm L}_{\rm obj}(\bm{q})$ of biased objects should be
given by a good model of nonlinear structure formation, such as the
halo model.

In this paper, we assume locality of the bias in Lagrangian space: the
Lagrangian density field $\rho^{\rm L}_{\rm obj}(\bm{q})$ is assumed
to be a function of a smoothed linear overdensity at the same
Lagrangian position,
\begin{equation}
  \delta_{R}(\bm{q}) = \int d^3q'\,W_{R}(|\bm{q} - \bm{q}'|)\,
  \delta_{\rm L}(\bm{q}'),
\label{eq:1-5}
\end{equation}
where $W_R$ is a smoothing kernel of size $R$, and $\delta_{\rm
  L}(\bm{q})$ is the (unsmoothed) linear overdensity. We call such
biasing scheme a ``local Lagrangian bias'' in this paper. As we will
show in the next section, the halo bias \cite{MW96} is actually a
special case of local Lagrangian bias. Another example of the
local Lagrangian bias is the peak bias with the approximation of
peak-background split \cite{BBKS,CK89,MJW97}. The locality of bias in
Lagrangian space does not mean locality in Eulerian space, because of
evolutionary effects. Therefore our biasing scheme does not fall into
the category of the local biasing scheme in a usual context of EPT.
The local Lagrangian bias is nonlocal in Eulerian space.

Thus, we introduce a Lagrangian bias function $F(\delta)$ by
\begin{equation}
  \rho^{\rm L}_{\rm obj}(\bm{q}) = 
  \bar{\rho}_{\rm obj}\, F[\delta_R(\bm{q})],
\label{eq:1-6}
\end{equation}
where $\bar{\rho}_{\rm obj}$ is the comoving mean density of the
biased objects, which is common in Lagrangian space and in Eulerian
space. This function has the following property:
\begin{equation}
  \left\langle F(\delta_R) \right\rangle = 1.
\label{eq:1-6-1}
\end{equation}
Equation (\ref{eq:1-3}) is equivalent to the following equation:
\begin{equation}
  \rho^{\rm E}_{\rm obj}(\bm{x}) =
  \bar{\rho}_{\rm obj}
  \int d^3q\, F[\delta_R(\bm{q})]\,
  \delta^3_{\rm D}[\bm{x} - \bm{q} - \bm{\Psi}(\bm{q})],
\label{eq:1-7}
\end{equation}
where $\delta^3_{\rm D}$ is 3-dimensional Dirac's delta function, and
we suppress the time-dependence for notational simplicity. Using the
Fourier transform of this equation, we obtain an expression of the
power spectrum of biased objects in Eulerian space,
\begin{multline}
  P_{\rm obj}(\bm{k}) = 
  \int d^3q\, e^{-i\bm{k}\cdot\bm{q}}
  \Biggl[
  \int \frac{d\lambda_1}{2\pi} \frac{d\lambda_2}{2\pi}
     \tilde{F}(\lambda_1) \tilde{F}(\lambda_2)
\\
     \times
     \left\langle
       e^{i[
         \lambda_1\delta_R(\bm{q}_1)
         + \lambda_2\delta_R(\bm{q}_2) ]
         -i\bm{k}\cdot[
         \bm{\Psi}(\bm{q}_1) - \bm{\Psi}(\bm{q}_2) ]
       }
     \right\rangle
    - 1
   \Biggr],
\label{eq:1-8}
\end{multline}
where $\tilde{F}(\lambda)$ is the Fourier transform of $F(\delta)$ and
$\bm{q} = \bm{q}_1 - \bm{q}_2$. The quantity in the ensemble average
$\langle\cdots\rangle$ in the above equation is a function of only
$\bm{q}$ because of translational invariance. In the absence of bias,
$F=1$ and $\tilde{F}(\lambda) = 2\pi \delta_{\rm D}(\lambda)$, the
Eq.~(\ref{eq:1-8}) reduces to a known expression \cite{PSlag}. We do
not assume rotational invariance for allowing our analysis to include
redshift-space clustering. Our convention of the power spectrum is
given by
\begin{equation}
  \left\langle
    \tilde{\delta}_{\rm obj}(\bm{k})
    \tilde{\delta}_{\rm obj}(\bm{k}')
  \right\rangle =
  (2\pi)^3 \delta^3_{\rm D}(\bm{k} + \bm{k}') P_{\rm obj}(\bm{k}),
\label{eq:1-9}
\end{equation}
where
\begin{equation}
  \tilde{\delta}_{\rm obj}(\bm{k}) =
  \int d^3x\, e^{-i\bm{k}\cdot\bm{x}}
  \left[
    \frac{\rho^{\rm E}_{\rm obj}(\bm{x})}{\bar{\rho}_{\rm obj}} - 1
  \right].
\label{eq:1-10}
\end{equation}
Similarly, the linear power spectrum $P_{\rm L}(k)$ is defined by a
similar equation to Eq.~(\ref{eq:1-9}) and a Fourier transform of
$\delta_{\rm L}(\bm{q})$.

The expression of Eq.~(\ref{eq:1-8}) has a form that we can apply to
the cumulant expansion theorem \cite{Ma85}
\begin{equation}
  \langle e^{-iX} \rangle
  = \exp\left[\sum_{N=1}^\infty \frac{(-i)^N}{N!} \langle X^N
      \rangle_{\rm c} \right],
\label{eq:1-11}
\end{equation}
where $\langle X^N \rangle_{\rm c}$ denotes a cumulant of a random
variable $X$ \cite{BCGS02}. The corresponding factor in
Eq.~(\ref{eq:1-8}) thus reduces to 
\begin{multline}
  \left\langle
    e^{i[
      \lambda_1\delta_R(\bm{q}_1)
      + \lambda_2\delta_R(\bm{q}_2) ]
      -i\bm{k}\cdot[
      \bm{\Psi}(\bm{q}_1) - \bm{\Psi}(\bm{q}_2) ]
    }
  \right\rangle \\
  =
  \exp\left[
    \sum_{n_1 + n_2 + m_1 + m_2 \geq 1}
    \frac{i^{n_1 + n_2 + m_1 + m_2}}
    {n_1!n_2!m_1!m_2!}
    {\lambda_1}^{n_1} {\lambda_2}^{n_2}
    B^{n_1 n_2}_{m_1 m_2}(\bm{k}, \bm{q})
  \right],
\label{eq:1-12}
\end{multline}
where the multinomial theorem is used, and
\begin{multline}
  B^{n_1 n_2}_{m_1 m_2}(\bm{k}, \bm{q})
  =
  (-1)^{m_1} \\ \times
  \left\langle
    [\delta_R(\bm{q}_1)]^{n_1} [\delta_R(\bm{q}_2)]^{n_2}
    [\bm{k}\cdot \bm{\Psi}(\bm{q}_1)]^{m_1}
    [\bm{k}\cdot \bm{\Psi}(\bm{q}_2)]^{m_2}
  \right\rangle_{\rm c}.
\label{eq:1-13}
\end{multline}
The translational invariance and the parity symmetry imply the
following identities:
\begin{align}
  B^{n_1 n_2}_{m_1 m_2}(\bm{k},\bm{q}) &=
  (-1)^{m_1 + m_2} B^{n_2 n_1}_{m_2 m_1}(\bm{k},-\bm{q}),
\label{eq:1-14a} \\
  &=
  (-1)^{m_1 + m_2} B^{n_1 n_2}_{m_1 m_2}(\bm{k},-\bm{q}),
\label{eq:1-14b}
\end{align}
and therefore Eq.~(\ref{eq:1-13}) is symmetric with respect to its
indices:
\begin{equation}
  B^{n_1 n_2}_{m_1 m_2}(\bm{k},\bm{q}) =
  B^{n_2 n_1}_{m_2 m_1}(\bm{k},\bm{q}).
\label{eq:1-15}
\end{equation}
When the initial density field is random Gaussian, which is assumed
throughout this paper, the Equation (\ref{eq:1-13}) of $m_1=m_2=0$
survives only when $n_1 + n_2 = 2$:
\begin{equation}
  B^{n_1 n_2}_{00}(\bm{k}, \bm{q})
  =
  \begin{cases}
    \xi_R(|\bm{q}|), & n_1 = n_2 = 1, \\
    \sigma_R^2, & (n_1 = 2, n_2 = 0)\ \mbox{or}\ (n_1 = 0, n_2 = 2), \\
    0, & \mbox{otherwise},
  \end{cases}
\label{eq:1-15-1}
\end{equation}
where $\sigma_R^2 = \xi_R(0)$, and $\xi_R(q)$ is the smoothed linear
correlation function of the linear density field,
\begin{equation}
  \xi_R(q) = \int \frac{k^2 dk}{2\pi^2} j_0(kq) W^2(kR)
  P_{\rm L}(k),
\label{eq:1-16}
\end{equation}
where $j_0(x) = x^{-1}\sin x$ is the spherical Bessel function of
zeroth order, and
\begin{equation}
    W(kR) = 4\pi \int x^2 dx j_0(kx) W_R(x)
\label{eq:1-16-1}
\end{equation}
is the window function of the smoothing kernel. For the equation
(\ref{eq:1-13}) of $n_1 = n_2 = 0$, we have
\begin{multline}
  B^{00}_{m_1 m_2}(\bm{k}, \bm{q}) \\
  =
  \begin{cases}
    A_{2m}(\bm{k}),
    & (m_1 = 2m, m_2 = 0)\ \mbox{or}\ (m_1 = 0, m_2 = 2m), \\
    B_{m_1 m_2}(\bm{k},\bm{q}),
    & m_1 \geq 1\ \mbox{and}\ m_2 \geq 1, \\
    0, & \mbox{otherwise},
  \end{cases}
\label{eq:1-17}
\end{multline}
where $m$ is a positive integer, and 
\begin{align}
  &A_{2m}(\bm{k}) \equiv
  \left\langle
    [\bm{k}\cdot \bm{\Psi}(\bm{0})]^{2m}
  \right\rangle_{\rm c},
\label{eq:1-18a}\\
  &B_{m_1 m_2}(\bm{k},\bm{q}) \equiv
  (-1)^{m_1}
  \left\langle
    [\bm{k}\cdot \bm{\Psi}(\bm{q}_1)]^{m_1}
    [\bm{k}\cdot \bm{\Psi}(\bm{q}_2)]^{m_2}
  \right\rangle_{\rm c}.
\label{eq:1-18b}
\end{align}

Using the above properties and quantities, and substituting
Eq.~(\ref{eq:1-12}) into Eq.~(\ref{eq:1-8}), we obtain an expression,
\begin{widetext}
\begin{multline}
  P_{\rm obj}(\bm{k}) = 
  \exp\left[
    2 \sum_{m=1}^\infty \frac{(-1)^m}{(2m)!}
    A_{2m}(\bm{k})
  \right]
  \int d^3q\, e^{-i\bm{k}\cdot\bm{q}}
  \exp\left[
    \sum_{m_1, m_2 \geq 1}^\infty
    \frac{ i^{m_1 + m_2}}{m_1!m_2!}
    B_{m_1 m_2}(\bm{k}, \bm{q})
  \right]
\\ \times
  \int_{-\infty}^\infty
  \frac{d\lambda_1}{2\pi} \frac{d\lambda_2}{2\pi}
  \tilde{F}(\lambda_1)
  \tilde{F}(\lambda_2)
  e^{-{\lambda_1}^2 {\sigma_R}^2/2 -{\lambda_2}^2 {\sigma_R}^2/2}
  \exp\left[
    - \lambda_1 \lambda_2 \xi_R(|\bm{q}|)
    + \sum_{n_1 + n_2\geq 1}^\infty \sum_{m_1 + m_2\geq 1}^\infty
    \frac{ i^{n_1 +n_2 + m_1 + m_2}}{n_1!n_1!m_1!m_2!}
    {\lambda_1}^{n_1} {\lambda_2}^{n_2}
    B^{n_1 n_2}_{m_1 m_2}(\bm{k}, \bm{q})
   \right]
\\ - (2\pi)^3 \delta_{\rm D}^3(\bm{k}).
\label{eq:1-19}
\end{multline}
\end{widetext}
So far the expression is formal and holds even on strongly nonlinear
scales. Gravitational nonlinear effects on the matter distribution and
on the bias are all included in the cumulants of Eq.~(\ref{eq:1-13}).
When the objects are unbiased, $F(\delta) = 1$, $\tilde{F}(\lambda) =
2\pi \delta_{\rm D}(\lambda)$, the second line of Eq.~(\ref{eq:1-19})
simply reduces to unity, and the expression is equivalent to the one
that was previously derived in Eq.~(8) of paper I \cite{Mat08}. In the
work of paper I, we applied the LPT to evaluate the cumulants of
Eq.~(\ref{eq:1-13}) when $n_1 = n_2 = 0$, and showed that expanding
only the exponential factor in the integrand and keeping the
exponential prefactor result in partial resummation of Eulerian
perturbations, which improves the standard EPT in quasi-linear regime.
This is justified by the fact that the exponential prefactor consists
only of cumulants of the displacement field at a single point, while
the remaining exponential factor consists of cumulants at two points
separated by $|\bm{q}| \sim |\bm{k}|^{-1}$. The latter cumulants are
small enough in a large-scale limit, $|\bm{k}| \rightarrow 0$. When
the exponential prefactor is expanded as well, this approach gives
equivalent results to the standard EPT. Evaluation of the cumulants in
redshift space is straightforward in the framework of LPT.

In the presence of Lagrangian biasing, a similar technique can be
adopted. In the integrand of $\lambda_1$ and $\lambda_2$, we expand
the last exponential factor, keeping the exponential prefactor which
involves $\sigma_R^2$. In a large-scale limit, $\sigma_R^2 \gg
\xi_R(|\bm{q}|)$ because $|\bm{q}|$ is large, and $\sigma_R^2 \gg
B^{n_1 n_2}_{m_1 m_2}$ ($m_1 + m_2 \geq 1$) because $|\bm{k}|$ is
small. Therefore, it is desirable to keep the exponential prefactor in
the integrand of $\lambda$'s. Expanding the last exponential factor,
the integrals of $\lambda$'s can be performed as
\begin{align}
  \int_{-\infty}^\infty \frac{d\lambda}{2\pi} \tilde{F}(\lambda)
  e^{- \lambda^2 \sigma_R^2/2} (i\lambda)^n &= 
  \frac{1}{\sqrt{2\pi}\, \sigma_R}
  \int_{-\infty}^{\infty} d\delta e^{-{\delta}^2/2{\sigma_R}^2}
  \frac{d^nF}{d{\delta}^n}
\nonumber\\
  &\equiv \left\langle F^{(n)} \right\rangle.
\label{eq:1-20}
\end{align}
The right-hand side (RHS) of Eq.~(\ref{eq:1-20}) corresponds to the
expectation value of derivatives of $F$. In the following, we use
notations such as $\langle F' \rangle$, $\langle F'' \rangle$ for $n =
1,2$, respectively, to represent the above integral. The local
Lagrangian bias is fully characterized by a series of these parameters
$\langle F^{(n)} \rangle$, which we call local Lagrangian bias
parameters.

The exponential prefactor in Eq.~(\ref{eq:1-19}) corresponds to the
characteristic function of the one-point distribution of displacement
field. In fact, from Eqs.~(\ref{eq:1-11}) and (\ref{eq:1-18a}) and
parity property, $A_{2m+1} = 0$, we have
\begin{equation}
  \exp\left[
    2 \sum_{m=1}^\infty \frac{(-1)^m}{(2m)!}
    A_{2m}(\bm{k})
  \right] =
  \left|\left\langle e^{-i\bm{k}\cdot\bm{\Psi}} \right\rangle\right|^2.
\label{eq:1-21}
\end{equation}
Evaluation of the above characteristic function of the RHS requires
the fully nonlinear dynamics. In this paper, the exponent of the LHS
is evaluated by adopting the perturbation theory.

Before closing this subsection, we comment on the role of smoothing
radius $R$. In the local Lagrangian biasing scheme with
peak-background split, the number density of biased objects is
spatially modulated by the linearly extrapolated background density
field in Lagrangian space. The large-scale clustering of biased
objects should not depend on the artificial choice of smoothing radius
$R$ to define the background field. In fact, in the case of halo bias,
it is explicitly shown that $\langle F^{(n)} \rangle$ is independent
on $R$. In our derivation, we have used an approximation $\sigma_R^2
\gg \xi_R(|\bm{q}|)$, which means our equations are valid on larger
scales than the smoothing radius of the Lagrangian bias: $k \ll
R^{-1}$. In particular, the smoothing kernel $W(kR)$ in final
expressions of our perturbation theory is replaced by unity, for
consistency with our approximation in the first place.

\subsection{
Biased power spectrum from Lagrangian perturbation
theory
\label{subsec:LPTeval}}

We evaluate the general Eq.~(\ref{eq:1-19}) via the LPT as outlined in
the previous section. Details of derivations by the one-loop
perturbation theory are given in the Appendix. In this subsection, we
summarize basic assumptions made in the derivations.

As explained above, we need to evaluate the cumulants
$B_{m_1m_2}^{n_1n_2}(\bm{k},\bm{q})$ of Eq.~(\ref{eq:1-13}). In LPT,
the displacement field is expanded as a perturbative series,
\begin{equation}
  \bm{\Psi} = \bm{\Psi}^{(1)} +  \bm{\Psi}^{(2)} +  \bm{\Psi}^{(3)} + 
  \cdots,
\label{eq:1-23}
\end{equation}
where $\bm{\Psi}^{(n)}$ is given by integrations over $n$th product
of linear density contrast $\delta_{\rm L}$ with kernels as in
Eq.~(\ref{eq:a-2}). By means of LPT, the cumulant of
Eq.~(\ref{eq:1-13}) reduces to an infinite sum over cumulants of
linear density contrast, which are straightforwardly given by the
linear power spectrum. On large scales where $|\bm{k}| \ll
|\bm{\Psi}|$, contributions from higher-order perturbations in
Eq.~(\ref{eq:1-23}) are small enough, and one can truncate the series.
In the perturbation theory, a consistent manner of truncation is given
by a loop-expansion \cite{BCGS02} to obtain a nonlinear power
spectrum. For Gaussian initial conditions, the loop expansion is
equivalent to the series expansion in terms of the linear power
spectrum $P_{\rm L}(k)$.

The treatment of nonlinear redshift-space distortions is simpler in
LPT than that in EPT as shown in paper I \cite{Mat08}. The
displacement field in redshift space $\bm{\Psi}^{\rm s}$ is given by
\begin{equation}
  \bm{\Psi}^{\rm s} = \bm{\Psi} + \frac{\hat{\bm {z}}\cdot
    \dot{\bm{\Psi}}}{H} \hat{\bm{z}},
\label{eq:1-24}
\end{equation}
where $\bm{\Psi}$ is the displacement field in real space, a dot
denotes the derivative with respect to the cosmic time $t$,
$\hat{\bm{z}}$ is a unit vector along the line of sight, $H =
\dot{a}/a$ is the time-dependent Hubble parameter, and $a(t)$ is the
scale factor. The relation between displacement fields in
Eq.~(\ref{eq:1-24}) is exactly linear even in the nonlinear regime. In
contrast, the redshift-space distortions of Eulerian variables are
given by nonlinear transformations. This is a reason why nonlinear
redshift-space distortions are easier to handle in LPT than that in
EPT.

The time dependence of each perturbative term in Eq.~(\ref{eq:1-23})
is approximately given by $\bm{\Psi}^{(n)} \propto D^n$, where $D(t)$
is the linear growth rate. This relation is exact in the
Einstein-de~Sitter model, and approximately holds in general cosmology
\cite{LPTapp,BCGS02}. We also apply the distant-observer approximation
in which the line of sight $\hat{\bm{z}}$ is fixed. The latter
approximation is commonly used in analyses of redshift-space
distortions and valid as long as the redshift surveys are deep enough
so that clustering scales of interest is smaller than distances
between the observer and galaxies \cite{Mat00}. With those
approximations, order-by-order linear transformations of displacement
fields become particularly simple: $\bm{\Psi}^{{\rm s}(n)} = R^{(n)}
\bm{\Psi}^{(n)}$, where $R^{(n)}$ is a $3\times 3$ matrix whose
components are given by Eq.~(\ref{eq:a-2-1}).

Keeping the one-loop LPT exact, the integrand of Eq.~(\ref{eq:1-19})
turns out to be a strongly oscillating function of $\bm{q}$. It seems
extremely difficult to numerically evaluate such integral. Instead, we
further expand and truncate the exponential factors in the integrand
at the one-loop level as explained in the previous subsection. This
means that our result is not exact at one-loop LPT, while the
neglected terms are of order ${\cal O}[P_{\rm L}(k)]^3$, which are
two- or or higher-loop contributions in terms of Eulerian
perturbations. Not expanding the exponential prefactor improves the
standard EPT as shown in paper I.

\subsection{
The linear power spectrum in real space and in redshift space
\label{subsec:LinPS}}

Expanding Eq.~(\ref{eq:1-19}) and keeping only linear terms in $P_{\rm
  L}(k)$, we obtain the biased power spectrum in linear perturbation
theory. The result in real space is simply given by a linear term of
Eq.~(\ref{eq:a-19}) with a substitution $f=0$, 
\begin{equation}
  P_{\rm obj}(k) =
  \left( 1 + \langle F' \rangle \right)^2
  P_{\rm L}(k).
\label{eq:2-1}
\end{equation}
Since the mass power spectrum is given by $P_{\rm m}(k) = P_{\rm
  L}(k)$ in linear theory, the linear bias factor $b$, which is
defined by
\begin{equation}
  P_{\rm obj}(k) = b^2P_{\rm m}(k),
\label{eq:2-2}
\end{equation}
is scale-independent:
\begin{equation}
  b = 1 + \langle F' \rangle,
\label{eq:2-3}
\end{equation}
i.e., $b$ does not depend on $k$ in a large-scale limit. In the
original halo approach, it is derived that the Eulerian linear bias
factor is given by the Lagrangian linear bias factor plus unity by
using a spherical collapse model \cite{MW96,MJW97,ST99} . The result
of Eq.~(\ref{eq:2-3}), which is derived without assuming spherical
collapse, is consistent to that approach. In this sense, the factor
$\langle F' \rangle$ corresponds to a Lagrangian linear bias factor.

It is interesting to notice that the linear bias should be
scale-independent in this limit, for any nonlinear function of $F$: in
linear perturbation theory, scale dependence cannot be produced by any
form of local Lagrangian bias. This result can be considered as a
generalization of the ``local bias theorem'' \cite{LocBias,SW98},
which states that the linear bias factor of local {\em Eulerian} bias
for sufficiently small $k$ is scale independent. The constancy of the
linear bias factor is now proven even for the local {\em Lagrangian}
bias, which is nonlocal in Eulerian space. It is known that the
additional constant term arises from small-scale inaccuracies of the
linear power spectrum, and the general asymptotic form of biased power
spectrum is given by $P_{\rm obj}(k) = b^2 P_{\rm m}(k) + c$ in a
large-scale limit \cite{SW98}.

The corresponding linear result in redshift space is given by a linear
term of Eq.~(\ref{eq:a-19}),
\begin{equation}
  P_{\rm obj}^{\rm (s)}(\bm{k}) = 
  \left(1 + \langle F' \rangle + f \mu^2 \right)^2
  P_{\rm L}(k),
\label{eq:2-4}
\end{equation}
where $\mu = \hat{\bm{z}}\cdot\bm{k}/k$ is the direction cosine of the
wavevector $\bm{k}$ with respect to the line of sight $\hat{\bm{z}}$,
$f = d\ln D/d\ln a = (HD)^{-1}\dot{D}$ is the logarithmic derivative
of the linear growth rate $D(t)$. This result is equivalent to the
Kaiser's formula \cite{Kaiser87}
\begin{equation}
  P_{\rm obj}^{\rm (s)}(\bm{k}) = 
  b^2 \left(1 + \beta \mu^2 \right)^2
  P_{\rm m}(k),
\label{eq:2-5}
\end{equation}
where the linear bias factor $b$ is given by Eq.~(\ref{eq:2-3}) and
$\beta = f/b$ is the redshift-space distortion parameter. Again, it is
interesting to notice that we have derived the Kaiser's formula in the
presence of any nonlinear local bias in Lagrangian space: the Kaiser's
formula with a scale-independent bias is a general consequence of a
large-scale limit even in this framework.

\subsection{One-loop corrections to the biased power spectrum
\label{subsec:OneloopGen}}

The formal expression of Eq.~(\ref{eq:1-19}) is evaluated by applying
the LPT. The derivation of one-loop corrections to the power spectrum
is detailed in the Appendix. The power spectrum in real
space with one-loop corrections is given by putting $f=0$ in
Eq.~(\ref{eq:a-19}). The result is
\begin{align}
  &P_{\rm obj}(k) =
  \exp\left[-\left(k/k_{\rm NL}\right)^2 \right]
\nonumber\\
  &\quad \times
  \Biggl\{
  \left( 1 + \langle F' \rangle \right)^2
  P_{\rm L}(k)
  + \frac{9}{98} Q_1(k) + \frac37 Q_2(k) + \frac12 Q_3(k)
\nonumber\\
  &\qquad + \langle F'\rangle \left[ \frac67 Q_5(k) + 2 Q_7(k) \right]
  + \langle F''\rangle \left[ \frac37 Q_8(k) + Q_9(k) \right]
\nonumber\\
  &\qquad + \langle F'\rangle^2 \left[ Q_9(k) + Q_{11}(k) \right]
  + 2 \langle F'\rangle \langle F''\rangle Q_{12}(k)
\nonumber\\
  &\qquad + \frac12 \langle F''\rangle^2 Q_{13}(k)
 + \frac67 \left( 1 + \langle F' \rangle \right)^2
  \left[ R_1(k) + R_2(k) \right]
\nonumber\\
  &\qquad
  - \frac{8}{21} \left( 1 + \langle F' \rangle \right) R_1(k)
  \Biggr\},
\label{eq:2-6}
\end{align}
where 
\begin{equation}
  k_{\rm NL}
  = \left[\frac{1}{6\pi^2} \int dk P_{\rm L}(k)\right]^{-1/2},
\label{eq:2-7}
\end{equation}
and the functions $Q_n(k)$, $R_n(k)$ are given by
Eqs.~(\ref{eq:a-13a})--(\ref{eq:a-15b}), and are second order in
$P_{\rm L}(k)$. When the exponential prefactor is expanded and only
second order terms in $P_{\rm L}(k)$ are retained, we obtain an
expression of EPT without any resummation of higher-order
perturbations. In an unbiased case, $\langle F' \rangle = \langle F''
\rangle = 0$, the expression reduces to the result of the one-loop
perturbation theory of mass \cite{MSS, Mat08}.

The power spectrum of the biased objects in redshift space with
one-loop corrections is given in Eq.~(\ref{eq:a-19}):
\begin{align}
  &P^{\rm (s)}_{\rm obj}(\bm{k}) =
  \exp\left\{-\left[1 + f(f+2)\mu^2\right]
      \left(k/k_{\rm NL}\right)^2 \right\}
\nonumber\\
  &\quad \times \left[
  \left( 1 + \langle F' \rangle + f\mu^2 \right)^2
  P_{\rm L}(k)
  + \sum _{n,m} \mu^{2n} f^m E_{nm}(k)
  \right],
\label{eq:2-8}
\end{align}
where $E_{nm}(k)$ is given by Eqs.~(\ref{eq:a-21a})--(\ref{eq:a-21i}).
When the bias is not present, $\langle F' \rangle = \langle F''
\rangle = 0$, this result reduces to the one derived in paper I
\cite{Mat08}.

A cross power spectrum of differently biased objects is similarly
given. When the bias functions of these objects are $F_1$ and $F_2$,
the cross power spectrum is given by substitutions of
Eqs.~(\ref{eq:a-6-1a})--(\ref{eq:a-6-1e}), after expanding
Eq.~(\ref{eq:2-6}) or (\ref{eq:2-8}) in terms of $\langle F' \rangle$
and $\langle F'' \rangle$. The spherical average of Eq.~(\ref{eq:2-8})
can be obtained by using the following integral:
\begin{align}
&  \frac12 \int_{-1}^1 d\mu e^{-x \mu^2} \mu^{2n}
  = \frac12 x^{-n-1/2} \gamma\left(n+\frac12,x\right)
\nonumber\\
& \qquad\qquad\qquad
= (-1)^n \frac{\sqrt{\pi}}{2}
     \left(\frac{d}{dx}\right)^n
    \left[\frac{{\rm erf}(x^{1/2})}{x^{1/2}}\right],
\label{eq:2-9}
\end{align}
where $\gamma(a,x)$ is the lower incomplete gamma function, and ${\rm
  erf}$ is the error function normalized by ${\rm erf}(+\infty) = 1$.
The correlation function is obtained by numerically Fourier transforming
Eq.~(\ref{eq:2-8}). Spherically averaged correlation function is
simply given by
\begin{equation}
    \xi(r) =
    \int_0^\infty \frac{k^2 dk}{2\pi^2} j_0(kr) P(k),
\label{eq:2-10}
\end{equation}
where $P(k)$ is the spherically averaged power spectrum.

As described in paper I, the origin of the exponential prefactor
$\exp[-(k/k_{\rm NL})^2]$ is the nonlinear smearing effect by random
motions of mass elements. In redshift space, additional smearing
effect is present along the lines of sight. The latter effect is
similar to the nonlinear fingers-of-God effect \cite{RDist}. The form
of exponential prefactor coincides with the one which has been
phenomenologically introduced in previous work \cite{Sco04,ESW07} to
represent the smearing effects.

The one-loop corrections in Eqs.~(\ref{eq:2-6}) and (\ref{eq:2-8}) are
given by $k$-dependent functions, $Q_1(k),\ldots,Q_{13}(k)$,
$R_1(k),R_2(k)$, which are defined by
Eqs.~(\ref{eq:a-13a})--(\ref{eq:a-15b}) in the Appendix. All
of these functions but $Q_{13}(k)$ vanish in a large-scale limit $k
\rightarrow 0$. The function $Q_{13}(k)$ contributes only when
$\langle F''\rangle \ne 0$. Thus, the one-loop contributions are
present even in a large-scale limit through $Q_{13}(k)$ when $\langle
F''\rangle \ne 0$. This function turns out to be a convolution of the
power spectrum,
\begin{equation}
  Q_{13}(k) = 
  \int \frac{d^3p}{(2\pi)^3} P_{\rm L}(p) P_{\rm L}(|\bm{k}-\bm{p}|) =
  \int d^3x\, e^{-i\bm{k}\cdot\bm{x}} \left[\xi_{\rm L}(x)\right]^2,
\label{eq:2-11}
\end{equation}
where $\xi_{\rm L}(x)$ is the linear correlation function in real
space. In configuration space, $[\xi_{\rm L}(x)]^2 \ll \xi_{\rm L}(x)$
in a large-scale limit, $x \rightarrow \infty$. However, in Fourier
space, $Q_{13}(k)$ has a finite value in a large-scale limit,
\begin{equation}
  Q_{13}(k \rightarrow 0) = 
  \int \frac{d^3p}{(2\pi)^3} \left[P_{\rm L}(p)\right]^2 =
  \int d^3x\, \left[\xi_{\rm L}(x)\right]^2,
\label{eq:2-12}
\end{equation}
while $P_{\rm L}(k\rightarrow 0) = 0$ for cold dark matter (CDM)-like
power spectra. Therefore, the power spectrum on very large scales is
dominated by a constant contribution originated from nonlinear
clustering when the bias is present and $\langle F''\rangle \ne 0$.
For related discussion within a framework of local Eulerian bias, see
Ref.~\cite{SW98}.

\subsection{Nonequivalence between the local Lagrangian bias and the
  local Eulerian bias
  \label{subsec:inequiv}}

In our local Lagrangian biasing scheme, bias parameters are given by a
set of parameters, $\{\langle F^{(n)} \rangle\}$. In the local
Eulerian biasing scheme, on the other hand, bias parameters are given
by a set of parameters $\{b_n\}$ which are coefficients of a Taylor
expansion,
\begin{equation}
    \delta_{\rm obj} = \sum_{n=0}^\infty \frac{b_n}{n!} \delta^n,
\label{eq:2-13}
\end{equation}
where $\delta_{\rm obj}$ is the overdensity of objects and $\delta$ is
the evolved overdensity of mass at the same Eulerian position with
some smoothing filter. One may wonder if there are some relations
between these two sets of bias parameters. However, the two biasing
schemes are not equivalent to each other, and the two sets of bias
parameters are not expressible from one another in general. Therefore,
the expression of the biased power spectrum derived above is
essentially different from the one with local Eulerian bias previously
derived in literature. Below we clarify this situation in detail.

An essential difference between those two schemes is that the Eulerian
bias is applied to dynamically evolved density fields while the
Lagrangian bias is applied to initial density fields. Since the
dynamical evolution is generally nonlocal, those two local biasing
schemes are not equivalent to each other.

In the standard halo approach, however, the Eulerian bias parameters
$\{b_n\}$ are derived \cite{MJW97,CS02}, although the halo bias falls
into a category of local Lagrangian bias. One may wonder if the local
Lagrangian bias is actually equivalent to the local Eulerian bias from
this fact. However, the spherical collapse model is crucially assumed
in such a derivation. Since the dynamical evolution in a spherical
collapse model is locally determined, the local Lagrangian bias and
the local Eulerian bias have one-to-one correspondence in such a
special case. It is only when the dynamical evolutions are treated
approximately as local processes that both biasing schemes become
equivalent to each other.

Similarly, the linear dynamical evolution is locally determined, and
there is a relation between linear bias parameters of two schemes,
$b_1 = 1 + \langle F'\rangle$. This is the reason why the linear power
spectrum of Eq.~(\ref{eq:2-1}) or (\ref{eq:2-4}) is equivalent to that
with Eulerian linear bias. There are not such relations for
higher-order bias parameters in general.

Accordingly, our one-loop result cannot be obtained by just a
reparametrization or renormalization of a set of parameters $\{b_n\}$
in one-loop EPT with local Eulerian bias. The power spectrum with
local Eulerian bias has a strong dependence on an artificial smoothing
length, and has a divergent result in a limit of small smoothing
length for CDM-like power spectra \cite{HMV98,HaloPT}. The power
spectrum with local Lagrangian bias derived above does not have such a
strong dependence and one can safely take the limit. Therefore the two
power spectra have qualitatively different properties and are never
equivalent to each other. The strong dependence on smoothing length
with local Eulerian bias can be removed by a renormalization scheme of
McDonald \cite{McD06}. Even in this case, the resulting power spectrum
[Eq.~(16) of Ref.~\cite{McD06}] is not reachable by simple
reparameterizations of our Eq.~(\ref{eq:2-6}), and vice versa. For
example, McDonald's Eq.~(16) has a common factor of $b_1^2$, while the
factor $(1 + \langle F' \rangle)^2$ cannot be factorized out in our
Eq.~(\ref{eq:2-6}) with any reparameterization of $\langle F''
\rangle$.

\section{
The halo bias in perturbation theory
\label{sec:HaloLPT}}

\subsection{
The halo approach as a local Lagrangian biasing scheme
\label{subsec:HaloBias}}

The nonlinear power spectrum derived above depends on the local
Lagrangian bias only through $\langle F^{(n)} \rangle$ defined by
Eq.~(\ref{eq:1-20}). Up to one-loop corrections, we only need two
numbers, $\langle F' \rangle$ and $\langle F'' \rangle$. For a general
local Lagrangian bias, these numbers could be considered as parameters
which should be fitted by observations. Alternatively, those numbers
can be derived once a model of bias function $F(\delta)$ is specified.
In this section, we take the latter approach, considering the halo
bias model.

So far the most successful biasing model in nonlinear structure
formation is provided by the halo approach
\cite{MW96,MJW97,SL99,ST99,Sel00,PS00,MF00,SSHJ01,CS02}, which is
based on the extended PS theory \cite{PS74, ExtPS}. In the PS
formalism, the mass of halo is related to the Lagrangian radius $R$ of
spherical cell by $M = 4\pi \bar{\rho} R^3/3$, or $R =[M/(1.162 \times
10^{12} h^{-1} M_\odot \Omega_{\rm m})]^{1/3} h^{-1}{\rm Mpc}$, where
$M_\odot = 1.989 \times 10^{30}\,{\rm kg}$ is the solar mass, and
$\Omega_{\rm m}$ is the density parameter at the present time. The
variance of mass overdensity in the cell, as a function of mass scale,
which is linearly extrapolated to the present time $z=0$, is given by
\begin{equation}
    \sigma^2(M) =
    \int \frac{k^2 dk}{2\pi^2} W^2(kR) P_0(k),
\label{eq:3-1}
\end{equation}
where $W(x) = 3(\sin x - x \cos x)/x^3$ is the top-hat window
function, $P_0(k) = P_{\rm L}(k)/D^2$ is the linear power spectrum
extrapolated to the present time.

The
critical overdensity, which is required for spherical collapse at
redshift $z$, and is linearly extrapolated to the present time, is
given by
\begin{equation}
  \delta_{\rm c}(z) = \frac{\delta_{\rm c}(0)}{D(z)}
\label{eq:3-3}
\end{equation}
where $D(z)$ is the linear growth rate as a function of redshift $z$,
normalized as $D(0) = 1$. The critical overdensity at the present
time, $\delta_{\rm c}(0)$, is given by $\delta_{\rm c}(0) =
3(3\pi/2)^{2/3}/5 = 1.68647$ in the Einstein-de~Sitter model. In
general cosmology, $\delta_{\rm c}$ depends weakly on cosmological
parameters and redshift \cite{CritD}. It is still a good approximation
to use the constant value of the Einstein-de~Sitter universe in
general cosmological models.


According to the PS theory, the comoving number density of haloes with
mass between $M$ and $M+dM$, identified at redshift $z$, is given by
\begin{equation}
  n(M,z)dM = \frac{2\bar{\rho}}{M} g(\nu) \frac{d\nu}{\nu},
\label{eq:3-5}
\end{equation}
where $\nu = \delta_{\rm c}(z)/\sigma(M)$, $g(\nu) = (2\pi)^{-1/2}\nu
\exp(-\nu^2/2)$, and $\bar{\rho}$ is the comoving mean density of
mass. In literature, a multiplicity function $f(\tilde{\nu})$ defined
by $g(\nu) = \tilde{\nu} f(\tilde{\nu})/2$ and $\tilde{\nu} = \nu^2$
is frequently introduced. The PS mass function is improved by Sheth
and Tormen (ST) \cite{ST99} to give a better fit to that of haloes in
numerical simulations of CDM-type cosmologies. The ST mass function is
also given by Eq.~(\ref{eq:3-5}) with a modified function
\begin{equation}
  g(\nu) = \frac{A(p)}{\sqrt{2\pi}}
  \left[ 1 + \frac{1}{(q\nu^2)^p} \right]
  \sqrt{q}\, \nu e^{-q\nu^2/2},
\label{eq:3-7}
\end{equation}
where $A(p) = [ 1 + \pi^{-1/2} 2^{-p} \Gamma(1/2 - p) ]^{-1}$, and
$p=0.3$ and $q=0.707$ are numerically fitted parameters. The PS mass
function is also given by Eq.~(\ref{eq:3-7}) with $p = 0$ and $q = 1$.
The Eq.~(\ref{eq:3-7}) has the following normalization,
\begin{equation}
   \int_0^\infty g(\nu) \frac{d\nu}{\nu} = \frac12,
\label{eq:3-8}
\end{equation}
which is equivalent to $\int n(M,z)MdM = \bar{\rho}$.

The original PS theory is extended to give the number density of
haloes of mass $M_1$, identified at redshift $z_1$, in a region of
Lagrangian radius $R_0$ in which the linear overdensity extrapolated
to the present time is $\delta_0$:
\begin{equation}
  n(M_1,z_1|\delta_0,R_0)dM_1 =
  \frac{2\bar{\rho}}{M_1} g(\nu_{10}) \frac{d\nu_{10}}{\nu_{10}},
\label{eq:3-9}
\end{equation}
where
\begin{align}
   &\nu_{10} = \frac{\delta_1 - \delta_0}
   {\left(\sigma_1^2 - \sigma_0^2\right)^{1/2}}, \\
   &\delta_1 = \delta_{\rm c}(z_1),\ \ 
    \sigma_0 = \sigma(M_0), \ \ 
    \sigma_1 = \sigma(M_1)
\label{eq:3-10}
\end{align}
and $M_0 = 4\pi \bar{\rho}R_0^3/3$. The haloes of mass $M_1$ are
collapsed at $z_1$, while $M_0$ is assumed uncollapsed at $z=0$, and
thus we always have $\delta_1 > \delta_0$.

The conditional number density of Eq.~(\ref{eq:3-9}) is interpreted as
a large-scale spatial modulation of halo densities in Lagrangian
space. The number density of haloes is a function of the linearly
extrapolated overdensity with smoothing radius $R_0$. Therefore,
Eq.~(\ref{eq:3-9}) corresponds to the Lagrangian number density of
biased objects in Eq.~(\ref{eq:1-6}) for haloes of mass between $M_1$
and $M_1 + dM_1$. The bias function for the halo bias is given by
\begin{equation}
  F_{M_1}(\delta_0) =
  \frac{n(M_1,z_1|\delta_0,R_0)dM_1}{n(M_1,z_1)dM_1},
\label{eq:3-11}
\end{equation}
where the mass scale $M_1$ is explicitly denoted. Defining
$\nu_1 =\delta_1/\sigma_1$, we have
\begin{equation}
  \frac{d\nu_{10}/\nu_{10}}{d\nu_1/\nu_1} = 
  \frac{\sigma_1^2}{\sigma_1^2 - \sigma_0^2},
\label{eq:3-13}
\end{equation}
where $\nu_1$,  $\nu_{10}$,  $\sigma_1$,  $\sigma_0$ are considered as
functions of halo mass $M_1$. Combining Eqs.~(\ref{eq:3-5}),
(\ref{eq:3-9}), (\ref{eq:3-11}), and (\ref{eq:3-13}), the halo bias
function reduces to
\begin{equation}
    F_{M_1}(\delta_0) = 
    \frac{\sigma_1^2}{\sigma_1^2 - \sigma_0^2}
    \frac{g(\nu_{10})}{g(\nu_1)}.
\label{eq:3-14}
\end{equation}
Since the RHS of Eq.~(\ref{eq:3-14}) depends on $\delta_0$ only through
$\delta_1 - \delta_0$ of $\nu_{10}$, we have
\begin{equation}
  \left\langle
      \left(\frac{\partial}{\partial \delta_0}\right)^n g(\nu_{10})
  \right\rangle =
  \left(-\frac{\partial}{\partial \delta_1}\right)^n
  \left\langle
      g(\nu_{10})
  \right\rangle,
\label{eq:3-15}
\end{equation}
where the average $\langle\cdots\rangle$ is taken over distribution of
$\delta_0$. Assuming the initial density field is Gaussian, the
distribution of $\delta_0$ is also Gaussian with variance
$\sigma_0^2$. In the case of PS mass function, $p=0$ and $q=1$, a
Gaussian integral to give the average $\langle g(\nu_{10}) \rangle$
can be exactly performed:
\begin{align}
  \left\langle g(\nu_{10}) \right\rangle &=
  \frac{1}{\sqrt{2\pi \sigma_0^2}}
  \int_{-\infty}^\infty d\delta_0\,
  e^{-\delta_0^2/2\sigma_0^2} g\left(\nu_{10}\right)
\nonumber \\
  &= \frac{\sigma_1^2 - \sigma_0^2}{\sigma_1^2} g\left(\nu_1\right).
\label{eq:3-16}
\end{align}
In a general case of ST mass function, $p \ne 0$ and $q \ne 1$,
Eq.~(\ref{eq:3-16}) is shown in a limit $\sigma_0 \ll \sigma_1$, which
is a reasonable approximation when $M_0$ is much larger than $M_1$.

\begin{figure}
\begin{center}
\includegraphics[width=19pc]{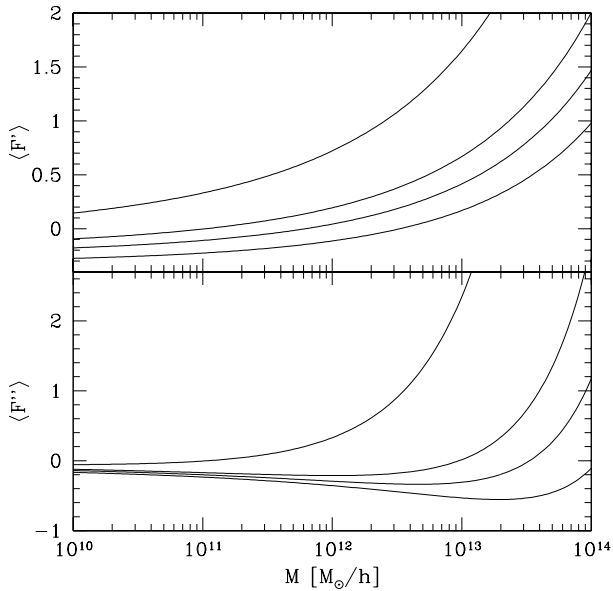}
\caption{\label{fig:fds} Local Lagrangian bias parameters $\langle
  F'\rangle$, $\langle F''\rangle$ as functions of halo mass.
  Different curves correspond to different redshifts ($z=0, 0.5, 1, 3$
  from bottom to top in each panel).}
\end{center}
\end{figure}

From Eqs.~(\ref{eq:3-14})--(\ref{eq:3-16}), we have
\begin{equation}
  \left\langle F_{M_1}^{(n)} \right\rangle =
  \frac{(-1)^n}{g(\nu_1)}
  \left(\frac{\partial}{\partial \delta_1}\right)^n g(\nu_1) =
  \frac{(-1)^n {\nu_1}^n}{{\delta_1}^n\, g(\nu_1)}
  \frac{\partial^n g(\nu_1)}{\partial {\nu_1}^n}.
\label{eq:3-17}
\end{equation}
With this expression, the following consistency relation is
straightforwardly shown:
\begin{equation}
  \frac{1}{\bar{\rho}}
  \int_0^\infty dM_1
    n(M_1,z_1) M_1
  \left\langle F_{M_1}^{(n)} \right\rangle =
  \begin{cases}
    1, & n=0, \\
    0, & n \geq 1.
  \end{cases}
\label{eq:3-18}
\end{equation}
For our purpose of one-loop corrections, we need only first two
derivatives. Dropping the subscript $M$ and '1', they are given by
\begin{align}
  &
  \left\langle F' \right\rangle =
  \frac{1}{\delta_{\rm c}(z)}
  \left[q \nu^2 - 1 + \frac{2p}{1 + \left(q\nu^2\right)^p}\right],
\label{eq:3-19a}\\
  &
  \left\langle F'' \right\rangle =
  \frac{1}{{\delta_{\rm c}}^2(z)}
  \left[q^2 \nu^4 - 3 q \nu^2
      + \frac{2p\left(2q\nu^2 + 2p - 1 \right)}
        {1 + \left(q\nu^2\right)^p}\right],
\label{eq:3-19b}
\end{align}
where $\nu = \delta_{\rm c}(z)/\sigma(M)$. The Eulerian bias factor of
Eq.~(\ref{eq:2-3}) with Eq.~(\ref{eq:3-19a}) agrees with that of the
original halo approach \cite{MW96,MJW97,ST99}. In Fig.~\ref{fig:fds},
$\langle F'\rangle$ and $\langle F''\rangle$ are plotted against mass
of haloes.
We adopt cosmological parameters $\Omega_{\rm m} = 0.28$,
$\Omega_{\Lambda} = 0.72$, $\Omega_{\rm b} = 0.046$, $h = 0.7$, $n_s =
0.96$ and $\sigma_8 = 0.82$ and the linear power spectrum $P_0(k)$ is
calculated from the output of the {\sc camb} code \cite{CAMB}. This
set of parameters is always assumed in the following figures
throughout this paper.

When a finite range of mass $[M_1,M_2]$ of haloes is considered, the
denominator and the numerator of Eq.~(\ref{eq:3-11}) are altered into
integrations in that range. The above derivation is similarly applied
in this case. As a result, we have
\begin{align}
  &
  \left\langle F^{(n)} \right\rangle =
  \frac{(-1)^n}{{\delta_{\rm c}}^n(z)}
  \frac{
    \displaystyle \int_{M_1}^{M_2} \nu^n
    \frac{d^n g}{d{\nu}^n} \frac{d\ln\sigma}{dM} \frac{dM}{M}
  }{
    \displaystyle \int_{M_1}^{M_2}
    g(\nu) \frac{d\ln\sigma}{dM} \frac{dM}{M}
  }.
\label{eq:3-20}
\end{align}
In a limit $M_2 \rightarrow M_1$, the Eq.~(\ref{eq:3-17}) is recovered
as expected.

\subsection{
Scale-dependence of the bias in quasi-linear regime
\label{subsec:HaloPSCF}}

Applying Eqs.~(\ref{eq:3-19a}) and (\ref{eq:3-19b}) to our results of
Eq.~(\ref{eq:2-6}) or Eq.~(\ref{eq:2-8}), a power spectrum of haloes
with one-loop corrections, either in real space or in redshift space,
can be evaluated. To demonstrate the effects of biasing and
redshift-space distortions, we show examples of power spectra and
correlation functions in this subsection, and briefly discuss the
impact on BAO features. Since the main purpose of the present paper is
to give a new formalism in perturbation theory, detailed investigation
of BAOs with our approach will be given elsewhere.

\begin{figure*}
\begin{center}
\includegraphics[width=30pc]{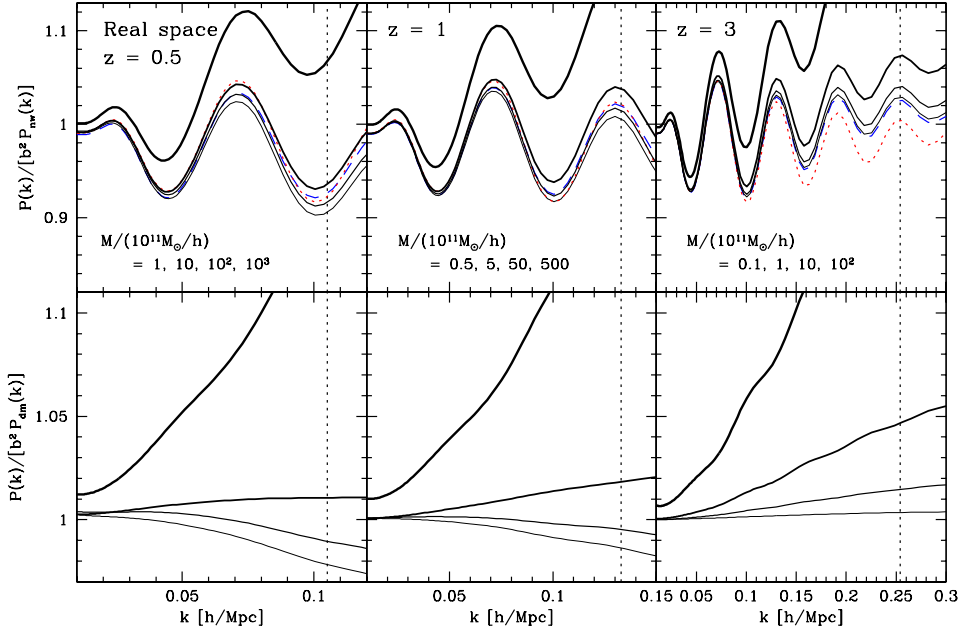}
\caption{\label{fig:realps} Dependencies on halo mass and redshift of
  nonlinear power spectrum in real space. In the top panels, each
  power spectrum is divided by a smoothed, no-wiggle linear power
  spectrum $P_{\rm nw}(k)$ \cite{EH99}, and by a squared linear bias
  factor $b^2$. Values of redshifts and halo masses are shown in each
  panel. Solid lines: nonlinear power spectra of haloes with different
  masses with increasing order from thinner to thicker lines; dotted
  lines: linear theory; dashed lines: nonlinear power spectra of dark
  matter. In the bottom panels, halo power spectra are divided by
  corresponding mass power spectra and by squared linear bias factor,
  presenting the scale dependence of halo bias. Vertical short-dashed
  lines correspond to the scale $k_{\rm NL}/2$ to indicate the
  validity range $k < k_{\rm NL}/2$, where our result is expected to
  be accurate within a few percent. }
\end{center}
\end{figure*}

\begin{figure*}
\begin{center}
\includegraphics[width=30pc]{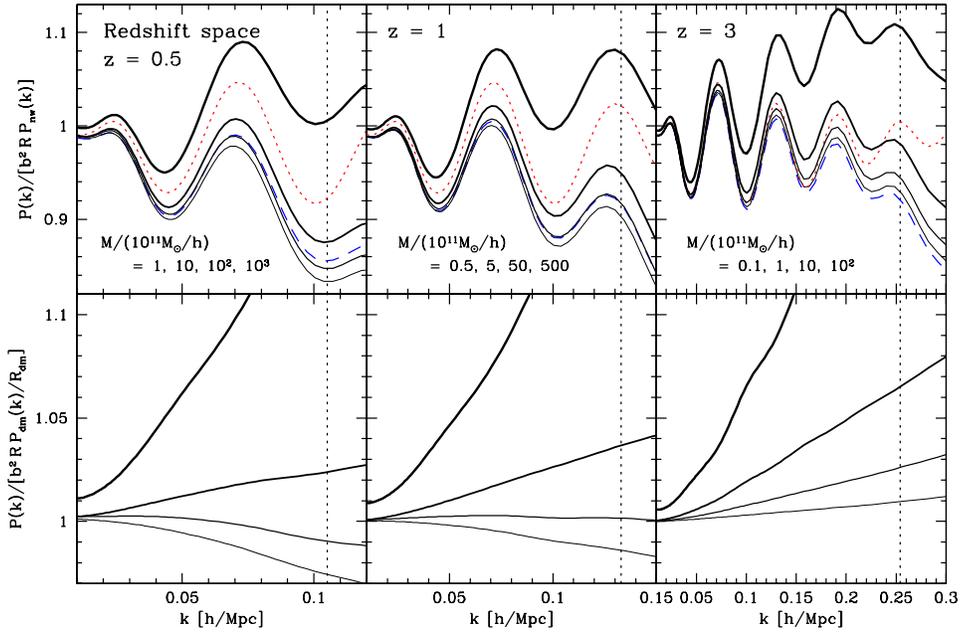}
\caption{\label{fig:redps} Same as Fig.~\ref{fig:realps}, but in
  redshift space. Spherically averaged power spectra are plotted.
  Linear redshift-space enhancement factor $R = 1 + 2\beta/3 +
  \beta^2/5$ is also scaled out. }
\end{center}
\end{figure*}

In Fig.~\ref{fig:realps}, the normalized power spectra in real space
are plotted, and those in redshift space are shown in
Fig.~\ref{fig:redps}.%
We adopt the same set of cosmological parameters as in
Fig.~\ref{fig:fds}. Angular averages are taken for power spectra in
redshift space. In paper I, it is shown that our one-loop formula is
valid within a few percent for $k < k_{\rm NL}/2$ \cite{Mat08},
compared to numerical simulations. Expecting this criterion is also
effective in our generalization including bias, the corresponding
scales of validity, $k_{\rm NL}/2$, are shown in vertical dotted lines
in the figures. To highlight nonlinear effects, overall amplitudes
predicted from linear growth rate, linear bias and linear
redshift-space distortions are scaled out. In real space, the
amplitude of power spectrum is proportional to a scale-independent
factor $D^2(z)b^2(z)$, where $b(z)$ is the linear bias factor defined
by Eq.~(\ref{eq:2-3}). In redshift space, Kaiser's enhancement factor
$R(z)=1 + 2\beta/3 + \beta^2/5$ \cite{Kaiser87} is an additional
source of the linear amplitude.

In upper panels in the figures, each power spectrum is normalized by a
smoothed, no-wiggle linear power spectrum $P_{\rm nw}(k)$ of
Ref.~\cite{EH99} to highlight baryonic features. The linear
amplification factors described above are all scaled out. In lower
panels, power spectra of haloes divided by those of dark matter are
plotted, where amplifications by linear biases and linear
redshift-space distortions are scaled out. Thus, curves in lower
panels show the scale dependence of bias. In the usual halo approach,
the scale dependence of the bias arises only from galaxy/dark matter
clustering within haloes, which is not considered in this paper. The
scale dependence shown in our results purely originates in clustering
of haloes themselves. Linear theory predicts constancy of halo bias on
large scales. Nonlinear effects of dynamics, biasing, and
redshift-space distortions are responsible for the scale dependence.

Comparing the clustering in real space and in redshift space, the
power spectra on small scales are suppressed by nonlinear
redshift-space distortions. This suppression is due to the large-scale
random motion of objects, which is similar to a phenomenon known as a
fingers-of-God effect \cite{RDist} on small scales. Generally, the
scale dependence of bias is strong for very massive and very light
haloes. Haloes of intermediate mass do not show significant deviations
from constant bias. The scale dependence of bias does not show
significant oscillations, and are mostly smooth functions of scales.
This is desirable for cosmological applications to use the BAO scale
as a standard ruler. Various nonlinear effects mostly modify the
broadband shape of power spectrum, and therefore resulting shifts of
acoustic scales are correctable as numerically demonstrated in
Ref.~\cite{SSEW08}.

\begin{figure*}
\begin{center}
\includegraphics[width=28pc]{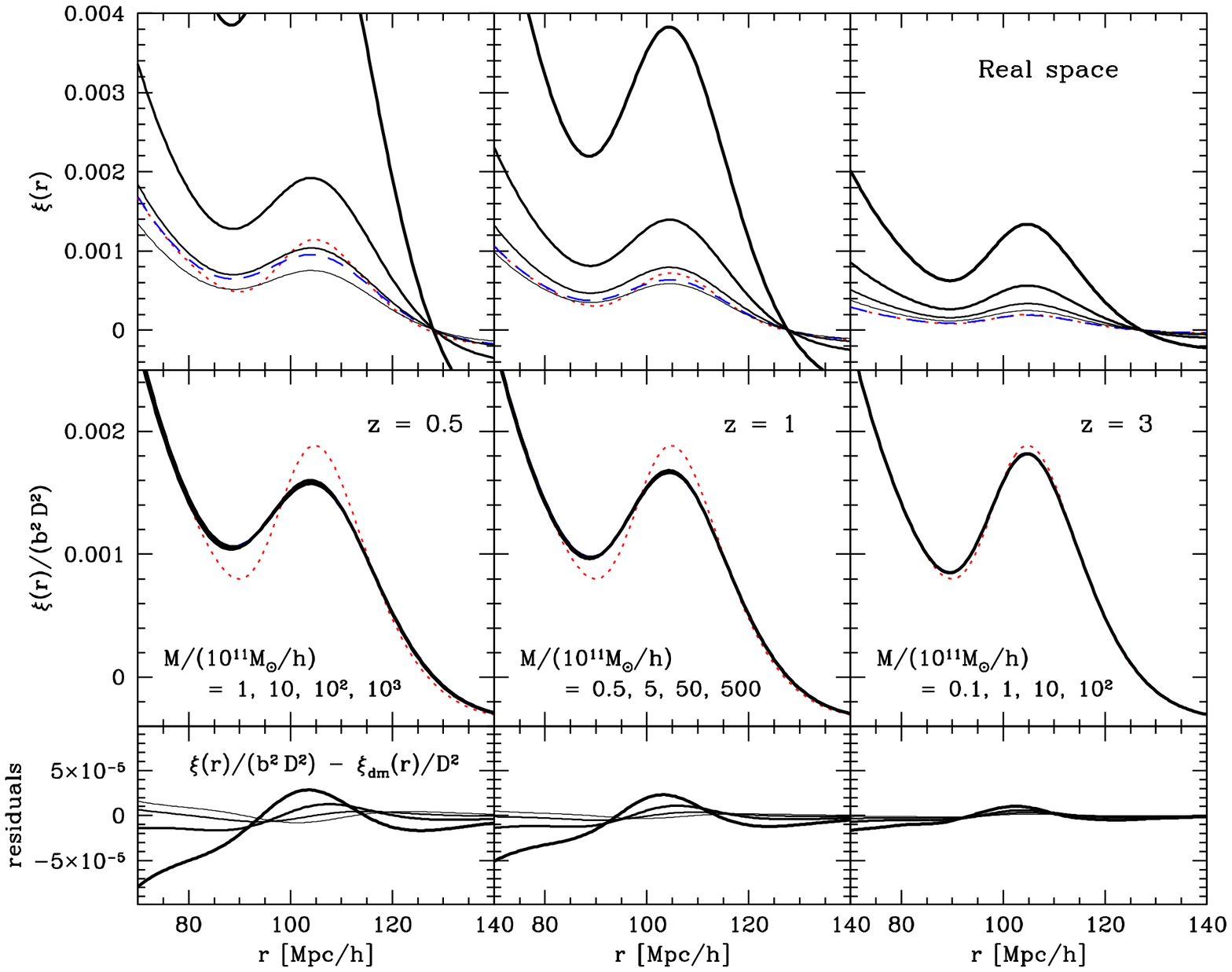}
\caption{\label{fig:realxi} Dependencies on halo mass and redshift of
  nonlinear correlation function in real space. Correlation functions
  with a fixed redshift and with different halo masses are presented
  in each column. Mass of the halo varies in increasing order from
  thinner to thicker solid lines. Dotted lines correspond to the
  prediction of linear theory and dashed lines correspond to nonlinear
  correlation functions of dark matter. In the top rows, the bare
  values of correlation function are plotted. In the middle rows, the
  correlation functions are normalized by linear bias factors and
  linear growth factors. In the bottom rows, residual values in the
  normalized correlation function of haloes (plotted in middle rows),
  relative to that of dark matter, are plotted. }
\end{center}
\end{figure*}

\begin{figure*}
\begin{center}
\includegraphics[width=28pc]{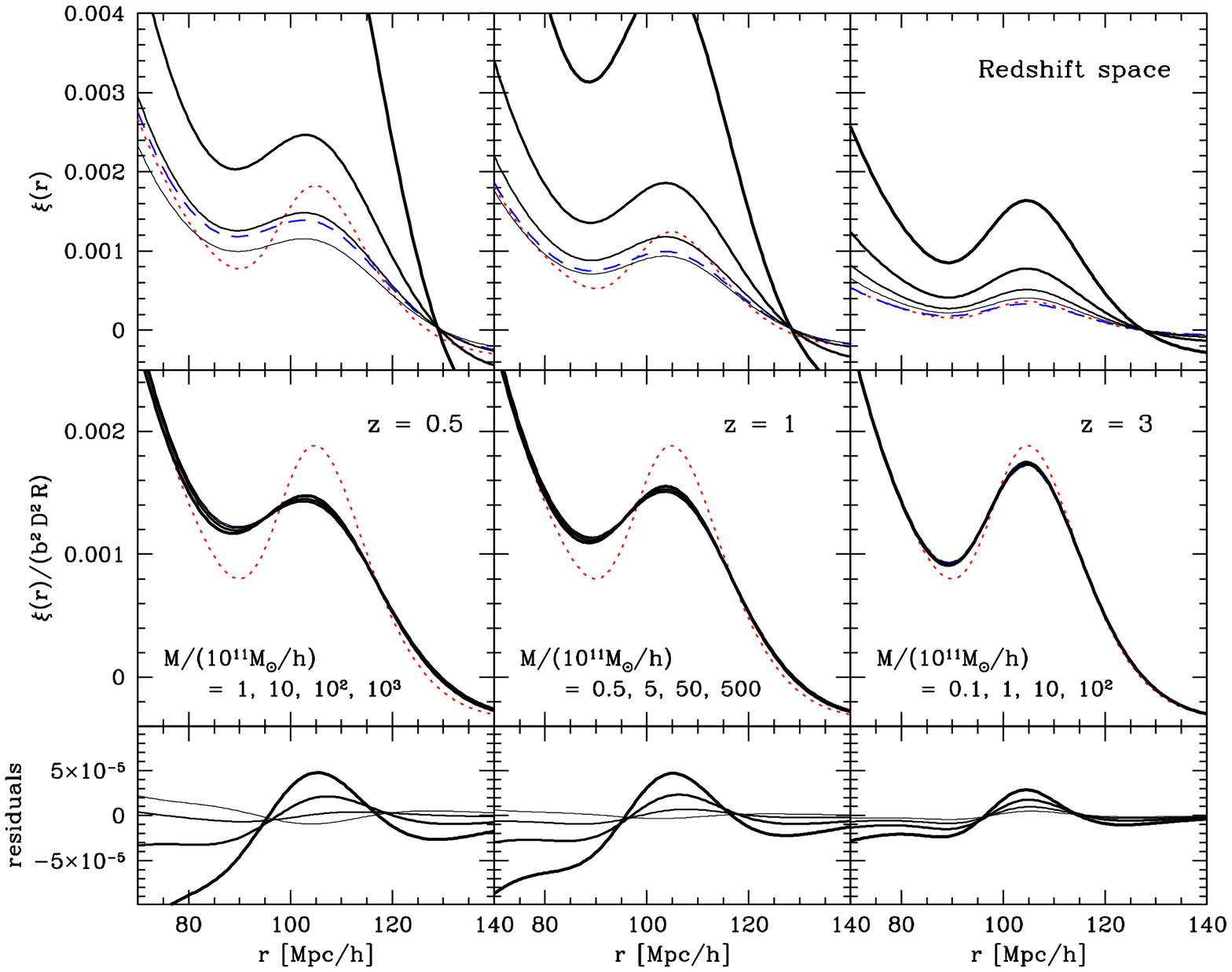}
\caption{\label{fig:redxi} Same as Fig.~\ref{fig:realxi}, but in
  redshift space. Spherically averaged correlation functions are
  plotted. Linear redshift-space enhancement factor $R$ is also scaled
  out. }
\end{center}
\end{figure*}

Although the power spectrum and correlation function are related by
Fourier transforms and have mathematically equivalent information,
cosmological information that can be extracted from them with real
data is not exactly equivalent to each other, because error properties
are different. While the BAO scales are imprinted in multiple wiggles
in the power spectrum, there appears one single peak in the
correlation function \cite{Mat04}. The physical BAO scale is just a
single scale, and many wiggles in the power spectrum are overtones of
the fundamental scale of BAOs. In Figs.~\ref{fig:realxi} and
\ref{fig:redxi}, correlation functions are plotted. They are
calculated by Fourier transforming the power spectra of
Figs.~\ref{fig:realps} and \ref{fig:redps}.
In paper I, we have shown that our method of one-loop corrections to
the clustering of dark matter, both in real space and in redshift
space, agrees very well with numerical simulations on scales $r >
70\,h^{-1}{\rm Mpc}$, where BAO signatures appear. Upper panels show
bare correlation functions, and middle panels show normalized
correlation functions. The overall amplitudes of the linear growth
rate, linear bias and linear redshift-space distortions are scaled out
in the normalized plots. The normalized correlation functions do not
significantly depend on mass of haloes, and plotted lines are quite
overlapped. In lower panels, residual values relative to normalized
correlation functions of dark matter are plotted.

Nonlinear effects degrade the signature of the BAO peak because of
random displacements of haloes, and the degradation is larger in
redshift space because of additional displacements along the lines of
sight. Halo bias does not significantly change the shape of the BAO
peak. This property is consistent with a recent analysis of numerical
simulations of halo clustering \cite{SBA08}. Effects of nonlinear
dynamics and nonlinear redshift-space distortions dominate those of
nonlinear halo bias. Massive haloes slightly enhance the BAO signature
by 5\% or so, while less massive haloes only change by 1--2\%. The
effects are slightly larger in redshift space.

\section{Summary
\label{sec:concl}
}

In this paper, we show that the nonlinear perturbation theory via the
Lagrangian picture is naturally incorporated with the local Lagrangian
bias on top of the redshift-space distortions. Explicit results of
one-loop power spectrum with a halo bias both in real space and in
redshift space are given. Our approach does not suffer from a
conceptual problem which appears in the EPT with local Eulerian bias.
The halo bias, as a local Lagrangian bias, is properly treated in the
Lagrangian picture of perturbations. Simultaneous inclusion of
redshift-space distortions is also natural in the Lagrangian picture.
Our approach involves a partial resummation of Eulerian perturbations,
and provides a better description in the quasilinear regime than the
standard EPT.

In our general framework of local Lagrangian bias, the bias is
characterized by local Lagrangian bias parameters $\langle F^{(n)}
\rangle$. We do not need a Taylor expansion of the bias function as
frequently adopted in local Eulerian biasing scheme. Only the first
two of the local Lagrangian bias parameters are needed in our one-loop
calculations. The results in Sec.~\ref{sec:LLBias} are applicable for
all biasing schemes as long as the bias is local in Lagrangian space.
The local Lagrangian bias parameters are considered as free parameters
if the bias model is not specified. Because of the nonlocal nature of
gravitational evolution, the local Lagrangian biasing scheme is not
equivalent to the local Eulerian biasing scheme.

There is a successful model of Lagrangian bias, i.e., the halo bias.
The general halo approach consists of several pieces of assumptions.
We adopt only the most fundamental concept of the halo approach, the
halo bias in Lagrangian space. The clustering of galaxies or dark
matter is dominated by halo clustering on large scales, where the
perturbation theory is safely applied. As a result, the local
Lagrangian bias parameters are calculated for halo bias without
ambiguity. The resulting power spectrum does not have any free
parameters once the mass or mass range of haloes is specified.

In usual linear analysis, the halo bias is independent on scales.
However, nonlinear effects introduce the scale dependence into the
halo bias. Such scale dependence could affect the determination of the
BAO scales unless the effect is properly quantified. Such effects are
quantitatively calculated without numerical simulations. We find that
the scale dependence of bias is a smoothly varying function.
Therefore, the BAO scales are shifted by change of the broadband shape
in the power spectrum. The correlation function is also affected by
scale dependence of nonlinear halo bias. Compared to the power
spectrum, the shape of the BAO peak in the correlation function seems
to be less affected. Those observations are consistent with the recent
results of numerical simulations \cite{Ang05,SBA08}.

Our formalism is compatible with any biasing scheme which is local in
Lagrangian space. For example, the peak bias is approximately
considered as a local in Lagrangian space in a limit of the
peak-background split \cite{BBKS,MJW97}, when the constraints imposed
by the spatial derivatives to define peaks can be neglected. However,
the last approximation may not be appropriate for accurately
predicting the BAO signature \cite{Des08}. The exact biasing mechanism
in the real world is definitely not local both in Eulerian space and
in Lagrangian space. The success of the halo approach indicates the
local Lagrangian bias is a good approximation at least on large
scales. However, extending the model of local Lagrangian bias to a
nonlocal one is an option to make the theory more accurate and
general. It is straightforward to extend our formalism in
Sec.~\ref{sec:LLBias} to include a nonlocal bias.

Although our formalism contains a partial resummation of higher-order
Eulerian perturbations, Lagrangian perturbations are truncated at the
one-loop level. Recent developments of the renormalized perturbation
theory and its variants \cite{RPTvar} show that it is possible to
reorganize and resum higher-order perturbations using the concept of
propagators. It would be interesting if one could use the concept of
propagators in Lagrangian space \cite{BV08} to further renormalize the
present formalism and to describe the deeply nonlinear regime, $k >
k_{\rm NL}/2$, which is not accessible by the present formalism.

We consider only the halo bias in this paper. The resulting power
spectrum corresponds to that of halo centers, or the two-halo term of
the halo model on large scales. In the halo approach, the nonlinear
power spectrum is given by a superposition of the one-halo term and
the two-halo term, which are given by convolutions with a model of
density profile of galaxies or dark matter. A model of density profile
is dominantly relevant to clustering on small scales. In a context of
the halo approach, our formalism improves the description of the
two-halo term, in which only the linear dynamics is usually included
because of simplicity. It is possible to include a model of density
profile of halo model in redshift space \cite{HaloRed}.

As pointed out in paper I, our approach does not have much power on
small scales in the power spectrum, $k > k_{\rm NL}/2$. The main
source of the powerless is the exponential damping factor, which
originates from random motion of the displacement field. A fully
nonlinear description of this factor may dramatically improve the
applicability of the present formalism for dark matter clustering.
However, in the presence of bias, the nonlinear regime is dominated by
the scale dependence of bias, which may be more appropriately
described by halo approach with the one-halo term, including a model
of density profile or halo occupation dynamics.

\begin{acknowledgments}
    I wish to thank A.~Taruya, R.~Takahashi for discussion. I
    acknowledge support from the Ministry of Education, Culture,
    Sports, Science, and Technology, Grant-in-Aid for Scientific
    Research (C), 18540260, 2006, and Grant-in-Aid for Scientific
    Research on Priority Areas No. 467 ``Probing the Dark Energy
    through an Extremely Wide and Deep Survey with Subaru Telescope.''
    This work is supported in part by JSPS (Japan Society for
    Promotion of Science) Core-to-Core Program ``International
    Research Network for Dark Energy.''
\end{acknowledgments}

\appendix

\section{
One-loop corrections to the biased power spectrum via the Lagrangian
perturbation theory
\label{app:evalpt}
}

In this Appendix, we outline a derivation of one-loop corrections to
the power spectrum with the local Lagrangian bias. Our goal is to
evaluate the cumulants of Eq.~(\ref{eq:1-13}) by the perturbation
theory, and obtain a perturbative expansion of Eq.~(\ref{eq:1-19}).
Most of the necessary techniques have been developed in paper I
\cite{Mat08}.

For the one-loop corrections, the cumulants of Eq.~(\ref{eq:1-13}) up
to second order in $P_{\rm L}(k)$ should be evaluated. We have applied
the LPT \cite{LPT, Mat08} to evaluate the similar cumulants in paper
I, and we can use the same method. In the LPT, the displacement field
$\bm{\Psi}$ is expanded by a perturbative series:
\begin{equation}
  \bm{\Psi} = \bm{\Psi}^{(1)} +  \bm{\Psi}^{(2)} +  \bm{\Psi}^{(3)} + 
  \cdots.
\label{eq:a-1}
\end{equation}
The first-order term $\bm{\Psi}^{(1)}$ corresponds to the classic
Zel'dovich approximation \cite{Zel70}. The spatial derivatives of each
term, $\partial_i \Psi_j^{(n)}$ have the order of ${\cal
  O}(\delta_{\rm L})^n$, where $\delta_{\rm L}$ is the linear density
field. In a Fourier representation,
\begin{multline}
 \tilde{\bm{\Psi}}^{(n)}(\bm{p}) =
  \frac{i}{n!}
  \int \frac{d^3p_1}{(2\pi)^3} \cdots \frac{d^3p_n}{(2\pi)^3}
  (2\pi)^3 \delta_{\rm D}^3\left(\sum_{j=1}^n \bm{p}_j - \bm{p}\right)
\\ \times
  \bm{L}^{(n)}(\bm{p}_1,\ldots,\bm{p}_n)
  \delta_{\rm L}(\bm{p}_1) \cdots \delta_{\rm L}(\bm{p}_n),
\label{eq:a-2}
\end{multline}
where $\delta_{\rm L}(\bm{p})$ is the Fourier transform of the linear
density field, and perturbative kernels $\bm{L}^{(n)}$ are given by
the LPT. Since the dependence of these kernels on time and on
cosmological parameters is weak, it is a good approximation to use
the kernels of the Einstein-de~Sitter model even in general cosmology
\cite{LPTapp}. In real space, expressions of the kernels
$\bm{L}^{(n)}$ up to third order are given by
\cite{Cat95,Mat08}
\begin{align}
& \bm{L}^{(1)}(\bm{p}_1) = \frac{\bm{k}}{k^2},
\label{eq:a-1-2a}\\
& \bm{L}^{(2)}(\bm{p}_1,\bm{p}_2)
  =\frac37 \frac{\bm{k}}{k^2}
  \left(1 - \mu_{1,2}^2\right),
\label{eq:a-1-2b}\\
& \bm{L}^{\rm (3a)}(\bm{p}_1,\bm{p}_2,\bm{p}_3)
  = \frac57 \frac{\bm{k}}{k^2}
  \left(1 - \mu_{1,2}^2\right)
  \left(1 - \mu_{12,3}^2 \right)
\nonumber\\
& \qquad
  - \frac13 \frac{\bm{k}}{k^2}
  \left(
      1 - 3\mu_{1,2}^2
      + 2 \mu_{1,2} \mu_{2,3} \mu_{3,1}
  \right)
  + \bm{k}\times \bm{T}(\bm{p}_1,\bm{p}_2,\bm{p}_3),
\label{eq:a-1-2c}
\end{align}
where $\bm{k} = \bm{p}_1 + \cdots + \bm{p}_n$ for each $\bm{L}^{(n)}$,
$\mu_{i,j} = \bm{p}_i \cdot \bm{p}_j/(p_i p_j)$, $\mu_{ij,k} =
(\bm{p}_i + \bm{p}_j) \cdot \bm{p}_k/(|\bm{p}_i + \bm{p}_j| p_k)$, and
a vector $\bm{T}$ represents a transverse part whose expression is not
needed in the following application. It is useful to symmetrize the
kernel $\bm{L}^{\rm (3a)}$ in terms of their arguments:
\begin{equation}
  \bm{L}^{(3)}(\bm{p}_1,\bm{p}_2,\bm{p}_3) =
  \frac13
  \left[\bm{L}^{\rm (3a)}(\bm{p}_1,\bm{p}_2,\bm{p}_3) + {\rm perm.}\right].
\label{eq:a-1-2d}
\end{equation}

As shown in paper I, the perturbative kernels in redshift space are
simply given by linear transformations by redshift-space distortion
tensors $R^{(n)}$, whose components are
\begin{equation}
  R^{(n)}_{ij} = \delta_{ij} + nf\hat{z}_i\hat{z}_j,
\label{eq:a-2-1}
\end{equation}
where $f = d\ln D/d\ln a = (HD)^{-1}\dot{D}$ is the logarithmic
derivative of linear growth rate $D(t)$ by the scale factor $a(t)$,
and $\hat{z}_i$ is a unit vector along the line of sight. In a matrix
notation, the kernels in redshift space is given by
\begin{equation}
  \bm{L}^{{\rm s}(n)} = R^{(n)}\bm{L}^{(n)}.
\label{eq:a-2-2}
\end{equation}

It is useful to define the following mixed polyspectra of linear
density field and displacement field:
\begin{multline}
  \left\langle
      \tilde{\delta}_{\rm L}(\bm{k}_1) \cdots
      \tilde{\delta}_{\rm L}(\bm{k}_l)
      \tilde{\Psi}_{i_1}(\bm{p}_1) \cdots
      \tilde{\Psi}_{i_m}(\bm{p}_m)
  \right\rangle_{\rm c} \\
  = (2\pi)^3 \delta_{\rm D}^3
  \left(\bm{k}_1 + \cdots + \bm{k}_l
      + \bm{p}_1 + \cdots + \bm{p}_m\right)\\
  \times
  (-i)^m
  C_{i_1\cdots i_m}
  \left(\bm{k}_1,\ldots,\bm{k}_l;\bm{p}_1,\ldots,\bm{p}_m\right),
\label{eq:a-3}
\end{multline}
where $\tilde{\delta}_{\rm L}$ and $\tilde{\Psi}_i$ are the Fourier
transforms of the linear density field and the displacement field,
respectively. When $l=0$, the above polyspectra are equivalent to the
ones defined in Eq.~(11) of paper I, but we adopt an opposite sign in
this paper. For $l+m=2$ in the Eq.~(\ref{eq:a-3}), we also use
notations such as
\begin{equation}
  C(\bm{k}) = C(\bm{k},-\bm{k}),\ 
  C_i(\bm{k}) = C_i(\bm{k};-\bm{k}),\ 
  C_{ij}(\bm{k}) = C_{ij}(\bm{k},-\bm{k}).
\label{eq:a-4}
\end{equation}
When $m=0$, the above polyspectra of Eq.~(\ref{eq:a-3}) is nonzero
only when $l=2$ for a Gaussian initial condition, which is assumed
throughout this paper. The Eq.~(\ref{eq:1-13}) has the order ${\cal
  O}[P_{\rm L}(k)]^{l_1 + l_2 + m_1 + m_2 - 1}$, because of the
property of cumulants \cite{BCGS02}. Therefore, we only need to
consider $n_1 + n_2 + m_1 + m_2 \leq 3$ up to one-loop corrections.
Expanding the exponential factors in Eq.~(\ref{eq:1-19}), but the
first prefactor, and truncating third- or higher-order terms in
$P_{\rm L}(k)$, we obtain
\begin{multline}
  P_{\rm obj}(\bm{k}) =
  \exp\left[
      k_i k_j \int \frac{d^3p}{(2\pi)^3} C_{ij}(\bm{p})
  \right]\\
  \times
  \left[
      a_{00}(\bm{k})
      + \langle F' \rangle\, a_{10}(\bm{k})
      + \langle F'' \rangle\, a_{01}(\bm{k})
      + \langle F' \rangle^2\, a_{20}(\bm{k})
  \right.
\\
  \left.
      + \langle F' \rangle \langle F'' \rangle\, a_{11}(\bm{k})
      + \langle F'' \rangle^2\, a_{02}(\bm{k})
  \right],
\label{eq:a-5}
\end{multline}
where
\begin{multline}
  a_{00}(\bm{k}) =
  -\, k_i k_j C_{ij}(\bm{k})
  - k_i k_j k_k \int \frac{d^3p}{(2\pi)^3}
    C_{ijk}(\bm{k},-\bm{p},\bm{p}-\bm{k})
\\
  + \frac12 k_i k_j k_k k_l
  \int \frac{d^3p}{(2\pi)^3}
    C_{ij}(\bm{p}) C_{kl}(\bm{k}-\bm{p}),
\label{eq:a-6a}
\end{multline}
\begin{multline}
  a_{10}(\bm{k}) =
  2 k_i C_i(\bm{k})
\\
   \hspace{1.5pc} + k_i k_j \int \frac{d^3p}{(2\pi)^3}
    \left[C_{ij}(\bm{k};-\bm{p},\bm{p}-\bm{k})
        -2 C_{ij}(-\bm{p};\bm{p}-\bm{k},\bm{k})
    \right]
\\
    -2 k_i k_j k_k \int \frac{d^3p}{(2\pi)^3}
     C_i(\bm{p}) C_{jk}(\bm{k}-\bm{p}),
\label{eq:a-6b}
\end{multline}
\begin{multline}
  a_{01}(\bm{k}) =
  -\, k_i \int \frac{d^3p}{(2\pi)^3} C_i(-\bm{p},\bm{p}-\bm{k};\bm{k})
\\
   + k_i k_j \int \frac{d^3p}{(2\pi)^3} C_i(\bm{p}) C_j(\bm{k}-\bm{p}),
\label{eq:a-6c}
\end{multline}
\begin{multline}
  a_{20}(\bm{k}) =
  C(\bm{k})
  + 2 k_i \int \frac{d^3p}{(2\pi)^3} C_i(\bm{k},-\bm{p};\bm{p}-\bm{k})
\\
  \hspace{2.5pc} + k_i k_j \int \frac{d^3p}{(2\pi)^3}
    \left[ C_i(\bm{p}) C_j(\bm{k}-\bm{p})
        -\, C(\bm{p}) C_{ij}(\bm{k}-\bm{p})
    \right],
\\
\label{eq:a-6d}
\end{multline}
\begin{equation}
  a_{11}(\bm{k}) =
  2 k_i \int \frac{d^3p}{(2\pi)^3} C(\bm{p}) C_i(\bm{k}-\bm{p}),
\label{eq:a-6e}
\end{equation}
\begin{equation}
  a_{02}(\bm{k}) =
  \frac12 \int \frac{d^3p}{(2\pi)^3} C(\bm{p}) C(\bm{k}-\bm{p}).
\label{eq:a-6f}
\end{equation}
We have neglected effects of the smoothing kernel $W(kR)$ in the above
equation, for the consistency of our treatment as discussed in the end
of Sec.~\ref{subsec:HaloBias}.
A cross power spectrum of differently biased objects is similarly
given. When the bias functions of these objects are $F_1$ and $F_2$,
the cross power spectrum is given by substitutions
\begin{align}
    \langle F' \rangle &\rightarrow
    \frac12
    \left(
        \langle F_1' \rangle + \langle F_2' \rangle
    \right),
\label{eq:a-6-1a}\\
    \langle F'' \rangle &\rightarrow
    \frac12
    \left(
        \langle F_1'' \rangle + \langle F_2'' \rangle
    \right),
\label{eq:a-6-1b}\\
    \langle F' \rangle^2 &\rightarrow
    \langle F_1' \rangle \langle F_2' \rangle,
\label{eq:a-6-1c}\\
    \langle F'' \rangle^2 &\rightarrow
    \langle F_1'' \rangle \langle F_2'' \rangle,
\label{eq:a-6-1d}\\
    \langle F' \rangle \langle F'' \rangle &\rightarrow
    \frac12
    \left(
        \langle F_1' \rangle\langle F_2'' \rangle
        + \langle F_1'' \rangle\langle F_2' \rangle
    \right),
\label{eq:a-6-1e}
\end{align}
in Eq.~(\ref{eq:a-5}).

Next we define mixed polyspectra of each order in perturbations:
\begin{multline}
  \left\langle
      \tilde{\delta}_{\rm L}(\bm{k}_1) \cdots
      \tilde{\delta}_{\rm L}(\bm{k}_l)
      \tilde{\Psi}^{(n_1)}_{i_1}(\bm{p}_1) \cdots
      \tilde{\Psi}^{(n_m)}_{i_m}(\bm{p}_m)
  \right\rangle_{\rm c} \\
  = (2\pi)^3 \delta_{\rm D}^3
  \left(\bm{k}_1 + \cdots + \bm{k}_l
      + \bm{p}_1 + \cdots + \bm{p}_m\right)\\
  \times
  (-i)^m
  C^{(n_1 \cdots n_m)}_{i_1\cdots i_m}
  \left(\bm{k}_1,\ldots,\bm{k}_l;\bm{p}_1,\ldots,\bm{p}_m\right),
\label{eq:a-7}
\end{multline}
where $\tilde{\Psi}^{(n)}_i$ are the Fourier transforms of the
displacement field of order $n$ in Eq.~(\ref{eq:a-1}). For $l+m=2$ in
the Eq.~(\ref{eq:a-7}), we also use notations similar to those in
Eq.~(\ref{eq:a-4}), such as $C^{(n)}(\bm{k})$, $C^{(n)}_i(\bm{k})$,
$C^{(n_1 n_2)}_{ij}(\bm{k})$. The Eq.~(\ref{eq:a-7}) is nonzero only
when $l + n_1 + \cdots + n_m$ is an even number. The original mixed
polyspectra of Eq.~(\ref{eq:a-3}) are given by sums of the polyspectra
of each order. In particular,
\begin{align}
  &C_i(\bm{p}) = C^{(1)}_i(\bm{p}) +  C^{(3)}_i(\bm{p}) + \cdots,
\label{eq:a-9b}\\
  &C_{ij}(\bm{p}) = C^{(11)}_{ij}(\bm{p}) +  C^{(22)}_{ij}(\bm{p})
  +  C^{(13)}_{ij}(\bm{p}) +  C^{(31)}_{ij}(\bm{p}) + \cdots,
\label{eq:a-9c}\\
  &C_i(\bm{p}_1,\bm{p}_2;\bm{p}_3) =
  C^{(2)}_i(\bm{p}_1,\bm{p}_2;\bm{p}_3) + \cdots,
\label{eq:a-9d}\\
  &C_{ij}(\bm{p}_1;\bm{p}_2,\bm{p}_3) =
  C^{(12)}_{ij}(\bm{p}_1;\bm{p}_2,\bm{p}_3) +
  C^{(21)}_{ij}(\bm{p}_1;\bm{p}_2,\bm{p}_3) + \cdots,
\label{eq:a-9e}\\
  &C_{ijk}(\bm{p}_1,\bm{p}_2,\bm{p}_3)
   = C^{(112)}_{ijk}(\bm{p}_1,\bm{p}_2,\bm{p}_3)
\nonumber\\
&\qquad
   + C^{(121)}_{ijk}(\bm{p}_1,\bm{p}_2,\bm{p}_3)
   + C^{(211)}_{ijk}(\bm{p}_1,\bm{p}_2,\bm{p}_3)
   + \cdots,
\label{eq:a-9f}
\end{align}
up to second order in $P_{\rm L}(k)$. Using the LPT kernels of
Eq.~(\ref{eq:a-2}), the mixed polyspectra of each order are given by
\begin{align}
  & C(\bm{p}) = P_{\rm L}(p),
\label{eq:a-10a}\\
  & C^{(1)}_i(\bm{p})
   = L^{(1)}_i(\bm{p}) P_{\rm L}(p),
\label{eq:a-10b}\\
  & C^{(11)}_{ij}(\bm{p})
   = - L^{(1)}_i(\bm{p}) L^{(1)}_j(\bm{p}) P_{\rm L}(p),
\label{eq:a-10c}
\end{align}
\begin{align}
  & C^{(3)}_i(\bm{p})
   = \frac12 P_{\rm L}(p) \int \frac{d^3p'}{(2\pi)^3}
     L^{(3)}_i(\bm{p},-\bm{p}',\bm{p}') P_{\rm L}(p'),
\label{eq:a-10d}\\
  & C^{(22)}_{ij}(\bm{p})
   = - \frac12 \int \frac{d^3p'}{(2\pi)^3}
   L^{(2)}_i(\bm{p}',\bm{p}-\bm{p}')
   L^{(2)}_j(\bm{p}',\bm{p}-\bm{p}')
\nonumber\\
 & \hspace{9pc} \times
   P_{\rm L}(p') P_{\rm L}(|\bm{p} - \bm{p}'|),
\label{eq:a-10e}\\
 &  C^{(13)}_{ij}(\bm{p}) = C^{(31)}_{ji}(\bm{p})
\nonumber\\
 & \hspace{0.5pc}  = - \frac12 L^{(1)}_i(\bm{p}) P_{\rm L}(p)
   \int \frac{d^3p'}{(2\pi)^3}
   L^{(3)}_j(\bm{p},-\bm{p}',\bm{p}') P_{\rm L}(p'),
\label{eq:a-10f}
\end{align}
\begin{align}
 & C^{(2)}_i(\bm{p}_1,\bm{p}_2;\bm{p}_3)
  = L^{(2)}_i(\bm{p}_1,\bm{p}_2) P_{\rm L}(p_1) P_{\rm L}(p_2),
\label{eq:a-11a}\\
 & C^{(12)}_{ij}(\bm{p}_1;\bm{p}_2,\bm{p}_3) =
  C^{(21)}_{ji}(\bm{p}_1;\bm{p}_3,\bm{p}_2)
\nonumber\\
 & \hspace{3pc}
   = - L^{(1)}_i(\bm{p}_2) L^{(2)}_j(\bm{p}_1,\bm{p}_2)
     P_{\rm L}(p_1) P_{\rm L}(p_2),
\label{eq:a-11b}\\
 &  C^{(112)}_{ijk}(\bm{p}_1,\bm{p}_2,\bm{p}_3)
   = C^{(211)}_{kij}(\bm{p}_3,\bm{p}_1,\bm{p}_2)
   = C^{(121)}_{jki}(\bm{p}_2,\bm{p}_3,\bm{p}_1)
\nonumber\\
 & \hspace{2pc}
   = L^{(1)}_i(\bm{p}_1) L^{(1)}_j(\bm{p}_2)
   L^{(2)}_k(\bm{p}_1,\bm{p}_2) P_{\rm L}(p_1) P_{\rm L}(p_2).
\label{eq:a-11c}
\end{align}

\begin{figure}
\includegraphics[width=15pc]{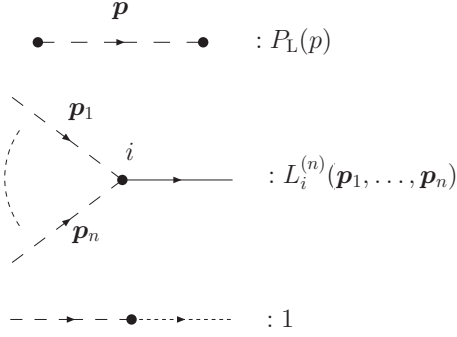}
\caption{\label{fig:feynrule} Feynman rules for diagrammatic
  representations. }
\end{figure}

\begin{figure}
\includegraphics[width=18pc]{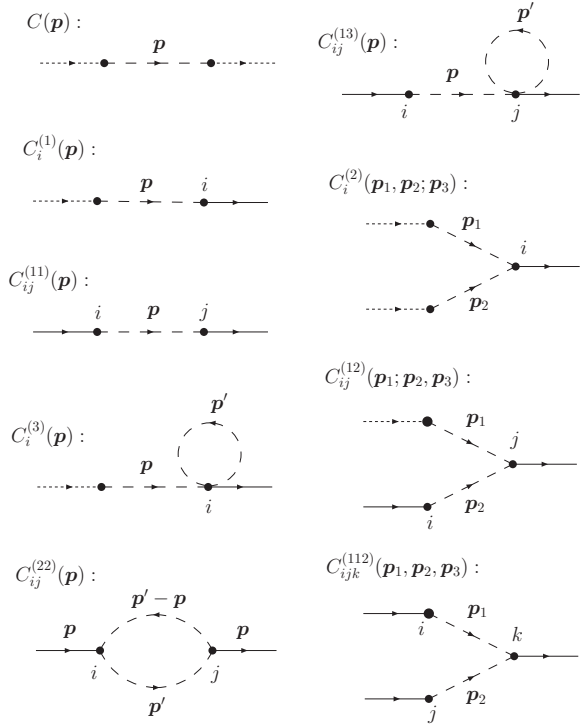}
\caption{\label{fig:feynpoly} Diagrammatic representations of
  polyspectra. }
\end{figure}

\begin{figure}
\includegraphics[width=17pc]{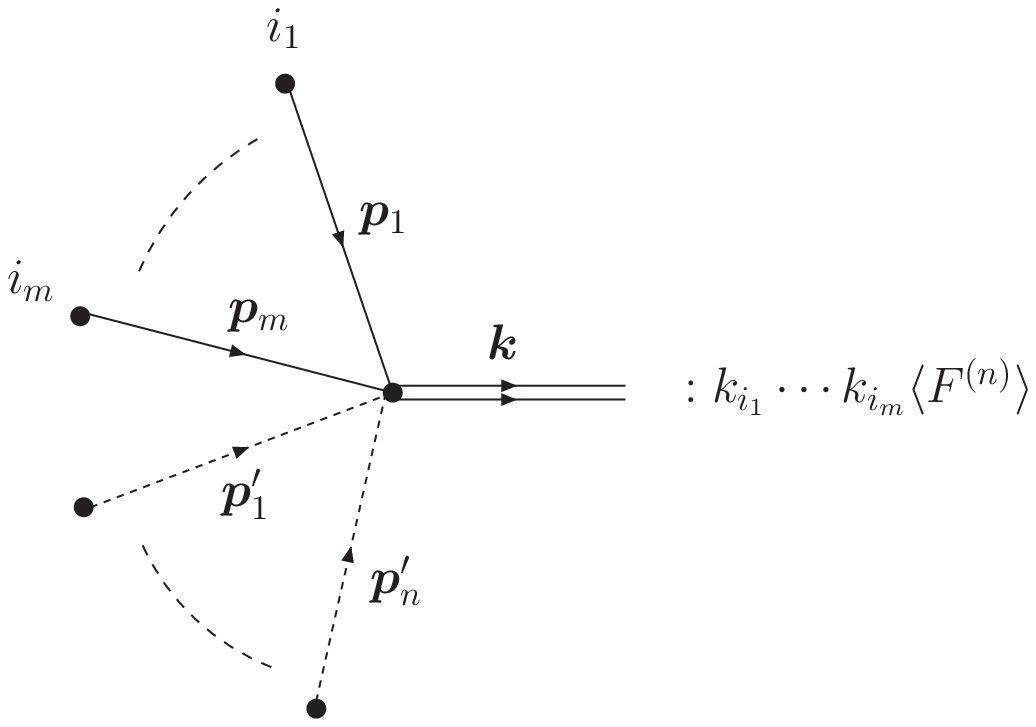}
\caption{\label{fig:feynext} Feynman rules for external lines. The
  momentum conservation, $\bm{k} = \bm{p}_1 + \cdots \bm{p}_m +
  \bm{p}_1' + \cdots \bm{p}_n'$, is assumed. }
\end{figure}

\begin{figure*}
\begin{center}
\includegraphics[width=42pc]{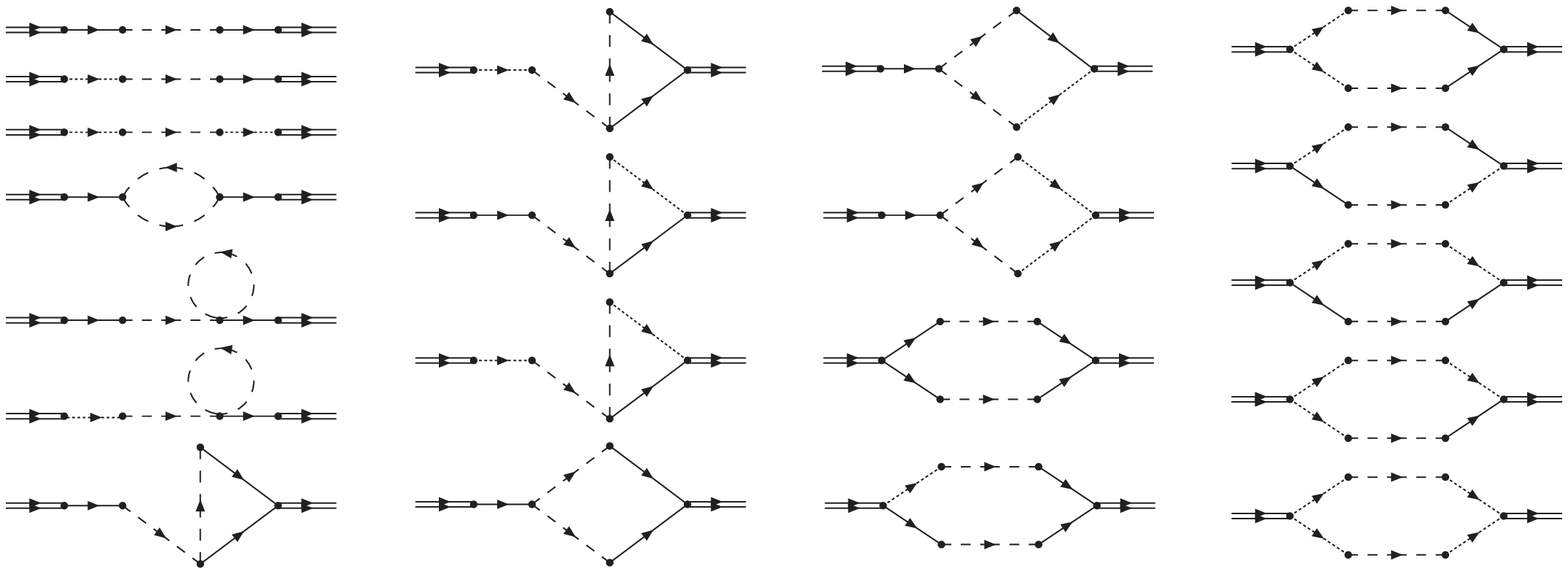}
\caption{\label{fig:feynps} All kinds of tree and one-loop diagrams
  for the biased power spectrum but an exponential prefactor.
  Topologically equivalent diagrams are not listed.}
\end{center}
\end{figure*}

\begin{figure}
\includegraphics[width=19pc]{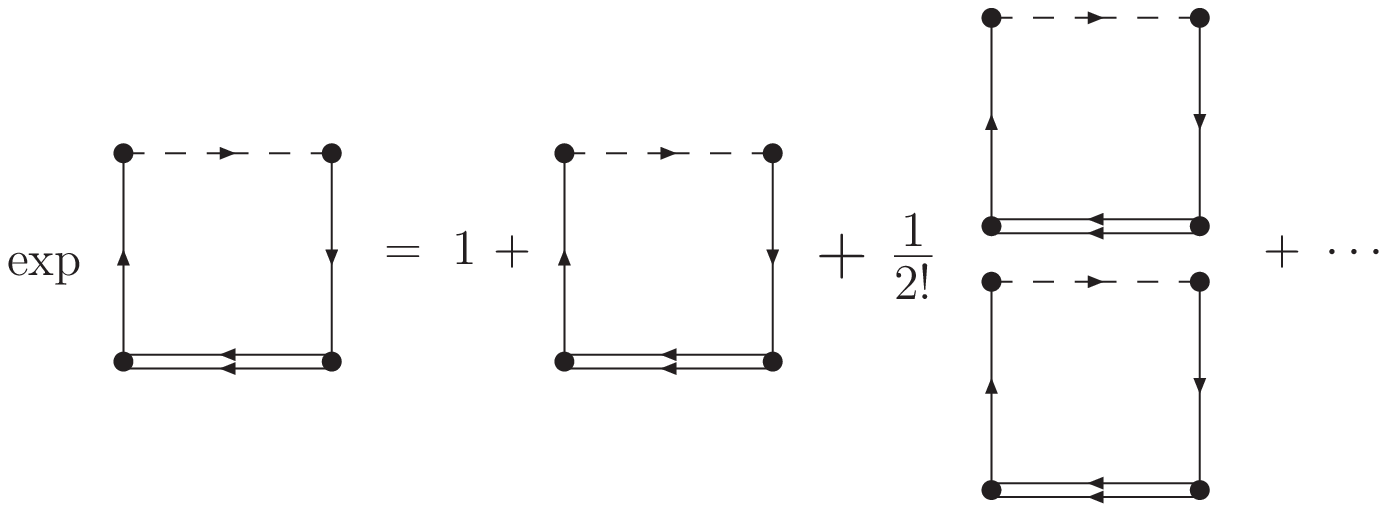}
\caption{\label{fig:feynexp} Diagrammatic representation of the
  exponential prefactor in one-loop approximation. }
\end{figure}

As in paper I, diagrammatic representations are helpful to understand
the structure of perturbative terms. With Feynman rules of
Fig.~\ref{fig:feynrule} and appropriate statistical factors,
Eqs.~(\ref{eq:a-10a})--(\ref{eq:a-11c}) are diagrammatically
represented in Fig.~\ref{fig:feynpoly}.
Substituting Eqs.~(\ref{eq:a-6a})--(\ref{eq:a-6f}) and
(\ref{eq:a-9b})--(\ref{eq:a-11c}) into Eq.~(\ref{eq:a-5}), we obtain a
lengthy expression of $P_{\rm obj}(k)$. For a diagrammatic
representation of the result, we introduce additional Feynman rules
for external lines in Fig.~\ref{fig:feynext}.
All the contributions to the power spectrum $P_{\rm obj}(k)$, but the
exponential prefactor, are diagrammatically given in
Fig.~\ref{fig:feynps}.
In the exponent of the exponential prefactor in Eq.~(\ref{eq:a-5}),
we consider only the first order in $P_{\rm L}(k)$, since the
remaining factor is already first order. In this approximation, the
exponential factor corresponds to the bubble diagrams in
Fig.~\ref{fig:feynexp}.
If we expand this exponential factor as well, we obtain a standard
loop expansion of the power spectrum. In the absence of bias, the
result agrees with that of standard EPT. In paper I, we pointed out
that keeping the exponential prefactor unexpanded provides a better
description of the nonlinear power spectrum in quasi-linear regime
than the standard EPT. In the diagrammatic interpretation, the use of
exponential prefactor corresponds to the partial renormalization of
the vacuum graphs, as depicted in Fig.~\ref{fig:feynexp}.

In real space, it is straightforward to evaluate Eq.~(\ref{eq:a-5})
using Eqs.~(\ref{eq:a-6a})--(\ref{eq:a-11c}) and perturbative kernels
in real space, $\bm{L}^{(n)}$. The above expressions are not confined
in real space, and they are also applicable for redshift-space
clustering when we use perturbative kernels $\bm{L}^{{\rm s}(n)}$ in
Eqs.~(\ref{eq:a-10a})--(\ref{eq:a-11c}). As a result, the mixed
polyspectra in redshift space are obtained from those in real space,
subjecting to linear transformations:
\begin{equation}
  C^{(n_1\cdots n_m)}_{i_1\cdots i_m} \rightarrow
  C^{{\rm s}(n_1\cdots n_m)}_{i_1\cdots i_m} =
  R^{(n_1)}_{i_1 j_1} \cdots R^{(n_m)}_{i_m j_m}
  C^{(n_1\cdots n_m)}_{j_1\cdots j_m}.
\label{eq:a-12}
\end{equation}
Therefore, it is desirable to first perform the integrations which
appear in Eqs.~(\ref{eq:a-6a})--(\ref{eq:a-6f}) with decomposed
polyspectra of Eqs.~(\ref{eq:a-9b})--(\ref{eq:a-9e}) in real space.
The calculations are similar to those presented in Appendix A of paper
I. Several integrations we need are already given there, and others
are not. To present the results, we first define the following
integrals:
\begin{align}
  Q_n(k) &=  \frac{k^3}{4\pi^2}
  \int_0^\infty dr\, P_{\rm L}(kr)
\nonumber\\
& \qquad \times
  \int_{-1}^1 dx
  P_{\rm L}[k(1 + r^2 - 2rx)^{1/2}]\,
  \tilde{Q}_n(r,x),
\label{eq:a-13a}\\
  R_n(k) &=
   \frac{k^3}{4\pi^2} P_{\rm L}(k)
    \int_0^\infty dr\, P_{\rm L}(kr)\,\tilde{R}_n(r),
\label{eq:a-13b}
\end{align}
where
\begin{equation}
  \tilde{Q}_1 = \frac{r^2 (1 - x^2)^2}{(1 + r^2 - 2rx)^2},
  \quad \tilde{Q}_2 = \frac{(1 - x^2) rx (1 - rx)}{(1 + r^2 - 2rx)^2},
\label{eq:a-14a}
\end{equation}
\begin{equation}
  \tilde{Q}_3 = \frac{x^2 (1 - rx)^2}{(1 + r^2 - 2rx)^2},
  \quad \tilde{Q}_4 = \frac{1 - x^2}{(1 + r^2 - 2rx)^2},
\label{eq:a-14b}
\end{equation}
\begin{equation}
  \tilde{Q}_5 = \frac{rx (1 - x^2)}{1 + r^2 - 2rx},
  \quad \tilde{Q}_6 = \frac{(1 - 3 rx)(1 - x^2)}{1 + r^2 - 2rx},
\label{eq:a-14c}
\end{equation}
\begin{equation}
  \tilde{Q}_7 = \frac{x^2 (1 - rx)}{1 + r^2 - 2rx},
  \quad \tilde{Q}_8 = \frac{r^2(1 - x^2)}{1 + r^2 - 2rx},
\label{eq:a-14d}
\end{equation}
\begin{equation}
  \tilde{Q}_9 = \frac{rx(1 - rx)}{1 + r^2 - 2rx},
  \quad \tilde{Q}_{10} = 1 - x^2,
\label{eq:a-14e}
\end{equation}
\begin{equation}
  \tilde{Q}_{11} = x^2,
  \quad \tilde{Q}_{12} = rx,
  \quad \tilde{Q}_{13} = r^2,
\label{eq:a-14g}
\end{equation}
and
\begin{align}
  \tilde{R}_1
  &= \int_{-1}^1 dx \frac{r^2 (1 - x^2)^2}{1 + r^2 - 2rx}
\nonumber\\
  &= - \frac{1 + r^2}{24 r^2}(3 - 14 r^2 + 3 r^4)
      + \frac{(r^2 - 1)^4}{16 r^3} \ln\left|\frac{1+r}{1-r}\right|,
\label{eq:a-15a}
\end{align}
\begin{align}
  \tilde{R}_2
  &= \int_{-1}^1 dx \frac{(1 - x^2) rx (1 - rx)}{1 + r^2 - 2rx}
\nonumber\\
  &= \frac{1 - r^2}{24 r^2}(3 - 2 r^2 + 3 r^4)
       + \frac{(r^2 - 1)^3 (1 + r^2)}{16 r^3}
       \ln\left|\frac{1+r}{1-r}\right|,
\label{eq:a-15b}
\end{align}
After lengthy algebra (see also paper I), we obtain
\begin{equation}
  C(\bm{k}) = P_{\rm L}(k),
\label{eq:a-16a}
\end{equation}
\begin{equation}
  C^{(1)}_i(\bm{k})
  = \frac{k_i}{k^2} P_{\rm L}(k),
\label{eq:a-16b}
\end{equation}
\begin{equation}
  C^{(3)}_i(\bm{k})
  = \frac{5}{21}\frac{k_i}{k^2} R_1(k),
\label{eq:a-16c}
\end{equation}
\begin{equation}
  C^{(11)}_{ij}(\bm{k})
  = -\frac{k_i k_j}{k^4} P_{\rm L}(k),
\label{eq:a-16d}
\end{equation}
\begin{equation}
  C^{(22)}_{ij}(\bm{k})
  = -\frac{9}{98} \frac{k_i k_j}{k^4} Q_1(k),
\label{eq:a-16e}
\end{equation}
\begin{equation}
  C^{(13)}_{ij}(\bm{k}) =   C^{(31)}_{ij}(\bm{k})
  = -\frac{5}{21} \frac{k_i k_j}{k^4} R_1(k),
\label{eq:a-16f}
\end{equation}
\begin{equation}
  \int \frac{d^3p}{(2\pi)^3}
   C^{(2)}_i(\bm{k},-\bm{p};\bm{p}-\bm{k})
   = \frac{3}{7} \frac{k_i}{k^2} \left[R_1(k) + R_2(k)\right],
\label{eq:a-17a}
\end{equation}
\begin{equation}
  \int \frac{d^3p}{(2\pi)^3}
   C^{(2)}_i(-\bm{p},\bm{p}-\bm{k};\bm{k})
   = - \frac{3}{7} \frac{k_i}{k^2} Q_8(k),
\label{eq:a-17b}
\end{equation}
\begin{align}
&  \int \frac{d^3p}{(2\pi)^3}
   C^{(12)}_{ij}(\bm{k};-\bm{p},\bm{p}-\bm{k})
   = \int \frac{d^3p}{(2\pi)^3}
   C^{(21)}_{ij}(\bm{k};-\bm{p},\bm{p}-\bm{k})
\nonumber\\
&\hspace{2.5pc}
   = - \frac{3}{14} \frac{\delta_{ij}}{k^2} R_1(k)
     + \frac{3}{14} \frac{k_i k_j}{k^4} [R_1(k) + 2R_2(k)],
\label{eq:a-17c}
\end{align}
\begin{equation}
  \int \frac{d^3p}{(2\pi)^3}
   C^{(12)}_{ij}(-\bm{p};\bm{p}-\bm{k},\bm{k})
   = - \frac{3}{7} \frac{k_i k_j}{k^4} Q_5(k),
\label{eq:a-17d}
\end{equation}
\begin{equation}
  \int \frac{d^3p}{(2\pi)^3}
   C^{(21)}_{ij}(-\bm{p};\bm{p}-\bm{k},\bm{k})
   = - \frac{3}{7} \frac{k_i k_j}{k^4} [R_1(k) + R_2(k)],
\label{eq:a-17e}
\end{equation}
\begin{align}
&  \int \frac{d^3p}{(2\pi)^3}
   C^{(112)}_{ijk}(\bm{k},-\bm{p},\bm{p}-\bm{k})
   = \int \frac{d^3p}{(2\pi)^3}
   C^{(121)}_{ijk}(\bm{k},-\bm{p},\bm{p}-\bm{k})
\nonumber\\
&\hspace{2pc}
   = \frac{3}{14} \frac{k_i \delta_{jk}}{k^4} R_1(k)
     - \frac{3}{14} \frac{k_i k_j k_k}{k^6} [R_1(k) + 2R_2(k)],
\label{eq:a-17f}
\end{align}
\begin{align}
&  \int \frac{d^3p}{(2\pi)^3}
   C^{(211)}_{ijk}(\bm{k},-\bm{p},\bm{p}-\bm{k})
\nonumber\\
&\hspace{2pc}
  = \frac{3}{14} \frac{k_i \delta_{jk}}{k^4} Q_1(k)
  - \frac{3}{14} \frac{k_i k_j k_k}{k^6} [Q_1(k) + 2Q_2(k)],
\label{eq:a-17g}
\end{align}
\begin{equation}
  \int \frac{d^3p}{(2\pi)^3} C(\bm{p}) C(\bm{k}-\bm{p})
  = Q_{13}(k),
\label{eq:a-18a}
\end{equation}
\begin{equation}
  \int \frac{d^3p}{(2\pi)^3} C(\bm{p}) C^{(1)}_i(\bm{k}-\bm{p})
  = \frac{k_i}{k^2} Q_{12}(k),
\label{eq:a-18b}
\end{equation}
\begin{align}
&  \int \frac{d^3p}{(2\pi)^3}
   C(\bm{p}) C^{(11)}_{ij}(\bm{k}-\bm{p})
\nonumber\\
&\hspace{2.5pc}
   = - \frac12 \frac{\delta_{ij}}{k^2} Q_{10}(k)
   + \frac12 \frac{k_i k_j}{k^4}[Q_{10}(k) - 2Q_{11}(k)],
\label{eq:a-18c}
\end{align}
\begin{align}
&  \int \frac{d^3p}{(2\pi)^3}
   C^{(1)}_i(\bm{p}) C^{(1)}_j(\bm{k}-\bm{p})
\nonumber\\
&\hspace{2.5pc}
   = - \frac12 \frac{\delta_{ij}}{k^2} Q_8(k)
   + \frac12 \frac{k_i k_j}{k^4}[Q_8(k) + 2Q_9(k)],
\label{eq:a-18d}
\end{align}
\begin{align}
&  \int \frac{d^3p}{(2\pi)^3}
   C^{(1)}_{(i}(\bm{p}) C^{(11)}_{jk)}(\bm{k}-\bm{p})
\nonumber\\
&\quad
  = -\frac12 \frac{\delta_{(ij} k_{k)}}{k^4} Q_6(k)
  + \frac12 \frac{k_i k_j k_k}{k^6}
     [Q_6(k) - 2Q_7(k)],
\label{eq:a-18e}
\end{align}
\begin{align}
&  \int \frac{d^3p}{(2\pi)^3}
   C^{(11)}_{(ij}(\bm{p}) C^{(11)}_{kl)}(\bm{k}-\bm{p})
  = \frac{3}{8}\frac{\delta_{(ij} \delta_{kl)}}{k^4} Q_1(k)
\nonumber\\
&\hspace{2.5pc}
  - \frac{1}{4}\frac{\delta_{(ij} k_k k_{l)}}{k^6}
     [3Q_1(k) + 12 Q_2(k) - 2 Q_4(k)]
\nonumber\\
&\hspace{2.5pc}
  + \frac{1}{8}\frac{k_i k_j k_k k_l}{k^8}
     [3Q_1(k) + 24Q_2(k) + 8Q_3(k) - 4Q_4(k)],
\label{eq:a-18f}
\end{align}
where the spatial indices are symmetrized over in
Eqs.~(\ref{eq:a-18e}) and (\ref{eq:a-18f}). In deriving the above
equations, the transverse part of Eq.~(\ref{eq:a-1-2c}) does not
contribute at all, because the rotational covariance implies
\begin{equation}
  \int \frac{d^3p}{(2\pi)^3}
  g(\bm{k},\bm{p}) \bm{T}(\bm{k},-\bm{p},\bm{p}) \propto \bm{k},
\label{eq:a-18-1}
\end{equation}
where $g$ is a scalar function.

Applying the transformation of Eq.~(\ref{eq:a-12}), the corresponding
integrals in redshift space are straightforwardly obtained. The
resulting integrals are substituted in
Eqs.~(\ref{eq:a-5})--(\ref{eq:a-6f}) with the help of
Eq.~(\ref{eq:a-9b})--(\ref{eq:a-9e}). The final result is represented
as
\begin{align}
  &P^{\rm (s)}_{\rm obj}(\bm{k}) =
  \exp\left\{-\left[1 + f(f+2)\mu^2\right]
      \left(k/k_{\rm NL}\right)^2 \right\}
\nonumber\\
  &\quad \times \left[
  \left( 1 + \langle F' \rangle + f\mu^2 \right)^2
  P_{\rm L}(k)
  + \sum _{n,m} \mu^{2n} f^m E_{nm}(k)
  \right],
\label{eq:a-19}
\end{align}
where 
\begin{equation}
  k_{\rm NL}
  = \left[\frac{1}{6\pi^2} \int dk P_{\rm L}(k)\right]^{-1/2},
\label{eq:a-20}
\end{equation}
and
\begin{align}
  E_{00} =& \frac{9}{98} Q_1 + \frac37 Q_2 + \frac12 Q_3 +
  \frac{10}{21} R_1 + \frac67 R_2 
\nonumber\\
  &+ \langle F' \rangle \left(
      \frac67 Q_5 + 2 Q_7 + \frac43 R_1 + \frac{12}{7} R_2 \right)
\nonumber\\
  &+ \langle F'' \rangle
  \left( \frac37 Q_8 + Q_9 \right)
  + \langle F' \rangle^2
  \left( Q_9 + Q_{11} + \frac67 R_1 + \frac67 R_2 \right)
\nonumber\\
  &+ 2 \langle F' \rangle\langle F'' \rangle Q_{12}
  + \frac12 \langle F'' \rangle^2 Q_{13},
\label{eq:a-21a}
\end{align}
\begin{align}
E_{11} =& \frac{18}{49} Q_1 + \frac{12}7 Q_2 + 2 Q_3 +
  \frac{40}{21} R_1 + \frac{24}{7} R_2
\nonumber\\
  &+ \langle F' \rangle \left(
      \frac{18}{7} Q_5 + 6 Q_7 + 4 R_1 + \frac{36}7 R_2 \right)
\nonumber\\
  &+ \langle F'' \rangle
  \left( \frac67 Q_8 + 2 Q_9 \right)
\nonumber\\
  &+ \langle F' \rangle^2
  \left( 2 Q_9 + 2 Q_{11} + \frac{12}{7} R_1 + \frac{12}{7} R_2 \right)
\nonumber\\
  &+ 2 \langle F' \rangle\langle F'' \rangle Q_{12},
\label{eq:a-21b}
\end{align}
\begin{align}
E_{12} =& -\frac{3}{14} Q_1 - \frac32 Q_2 + \frac14 Q_4 - \frac67 R_1
  + \langle F' \rangle \left( Q_6 - \frac67 R_1 \right)
\nonumber\\
  &- \frac12  \langle F'' \rangle Q_8
  - \frac12 \langle F' \rangle^2 \left( Q_8 - Q_{10} \right),
\label{eq:a-21c}
\end{align}
\begin{align}
E_{22} =& \frac{57}{98} Q_1 + \frac{51}{14} Q_2 + 3 Q_3
   - \frac14 Q_4 + \frac{16}{7} R_1 + \frac{30}{7} R_2
\nonumber\\
  &+ \langle F' \rangle \left(
      \frac{12}{7} Q_5 - Q_6 + 6 Q_7 + \frac{18}{7} R_1
      + \frac{24}{7} R_2 \right)
\nonumber\\
  & + \langle F'' \rangle
  \left( \frac12 Q_8 + Q_9 \right)
\nonumber\\
  &+ \langle F' \rangle^2
  \left( \frac12 Q_8 + Q_9 -\frac12 Q_{10} + Q_{11} \right),
\label{eq:a-21d}
\end{align}
\begin{equation}
E_{23} = - \frac37 Q_1 - 3 Q_2 + \frac12 Q_4 - \frac67 R_1
 + \langle F' \rangle Q_6,
\label{eq:a-21e}
\end{equation}
\begin{equation}
E_{24} = \frac{3}{16} Q_1,
\label{eq:a-21f}
\end{equation}
\begin{align}
E_{33} =& \frac37 Q_1 + \frac{27}{7} Q_2 + 2 Q_3 - \frac12 Q_4
   + \frac67 R_1 + \frac{12}{7} R_2 
\nonumber\\
   &+ \langle F' \rangle \left( - Q_6 + 2 Q_7 \right),
\label{eq:a-21g}
\end{align}
\begin{equation}
E_{34} = -\frac38 Q_1 - \frac32 Q_2 + \frac14 Q_4,
\label{eq:a-21h}
\end{equation}
\begin{equation}
E_{44} = \frac{3}{16} Q_1 + \frac32 Q_2 + \frac12 Q_3 - \frac14 Q_4,
\label{eq:a-21i}
\end{equation}
and all the other $E_{nm}$ which are not listed above are zero.

Equations (\ref{eq:a-19})--(\ref{eq:a-21i}), together with
Eqs.~(\ref{eq:a-13a})--(\ref{eq:a-15b}), are complete set of equations
to give the general one-loop power spectrum with effects of local
Lagrangian bias and redshift-space distortions. Although the number of
terms are large, they are all given by simple integrals of $Q_n(k)$
and $R_n(k)$ of Eqs.~(\ref{eq:a-13a}) and (\ref{eq:a-13b}), which are
numerically easy to evaluate. The power spectrum in real space is
obtained by simply putting $f=0$. When the bias is not present,
$\langle F' \rangle = \langle F'' \rangle = 0$, this result exactly
agrees with the one which is derived in paper I. A cross power
spectrum is obtained by substitutions of
Eqs.~(\ref{eq:a-6-1a})--(\ref{eq:a-6-1e}), after expanding
Eq.~(\ref{eq:a-19}) in terms of $\langle F' \rangle$ and $\langle F''
\rangle$.


\bigskip

\newcommand{\apjl}{Astrophys. J. Letters}
\newcommand{\apjs}{Astrophys. J. Suppl. Ser.}
\newcommand{\mnras}{Mon. Not. R. Astron. Soc.}
\newcommand{\pasj}{Publ. Astron. Soc. Japan}
\newcommand{\apss}{Astrophys. Space Sci.}
\newcommand{\aap}{Astron. Astrophys.}
\newcommand{\physrep}{Phys. Rep.}
\newcommand{\mpla}{Mod. Phys. Lett. A}
\newcommand{\jcap}{J. Cosmol. Astropart. Phys.}


\end{document}